\documentclass[aps,showpacs,preprint,groupedaddress,amsmath,amssymb,letterpaper,superscriptaddress]{revtex4}
\usepackage{graphicx}
\usepackage{dcolumn} 
\usepackage{bm}      
\usepackage{times}
\usepackage{color}

\usepackage{setspace}

\newcommand{\lansa}{{LANS$-\alpha$} }

\newcommand{\lans}{LANS$-\alpha$}
\newcommand{\Res}{Reynolds numbers}

\newcommand{\dof}{{\sl dof}}
\newcommand{\dofa}{{\sl dof} }
\newcommand{\pdf}{{\sl pdf}}
\newcommand{\pdfa}{{\sl pdf} }
\renewcommand{\vec}{\mathbf}
\newcommand{\uvec}{\bar{\vec{v}}}

\def\resp#1{#1}  
\newcommand{\neu}[2]{{#1}{#2}}
\newcommand{\add}[2]{{#1}{#2}}
\def\thesis#1{}                          
\def\paper#1{#1}                         

\begin{document}

\title{Highly turbulent solutions of \lansa and their LES potential}
 
 \author{Jonathan \surname{Pietarila Graham}}
 \affiliation{National
  Center for Atmospheric Research,\footnote{The National Center for
  Atmospheric Research is sponsored by the National Science
  Foundation} P.O. Box 3000, Boulder, Colorado 80307, USA}
 \affiliation{currently at Max-Planck-Institut f\"ur 
Sonnensystemforschung, 37191 Katlenburg-Lindau, Germany}
  \author{Darryl D. Holm}
  \affiliation{Department of
  Mathematics, Imperial College London, London SW7 2AZ, UK}
  \affiliation{Computer and Computational Science Division, Los Alamos National
  Laboratory, Los Alamos, NM 87545, USA} 
  \author{Pablo D. Mininni}
 \affiliation{National
  Center for Atmospheric Research,\footnote{The National Center for
  Atmospheric Research is sponsored by the National Science
  Foundation} P.O. Box 3000, Boulder, Colorado 80307, USA}
 \affiliation{Departamento 
de F\'\i sica, Facultad de Ciencias Exactas y Naturales, Universidad de 
Buenos Aires, Ciudad Universitaria, 1428 Buenos Aires, Argentina}
  \author{Annick Pouquet} 
 \affiliation{National
  Center for Atmospheric Research,\footnote{The National Center for
  Atmospheric Research is sponsored by the National Science
  Foundation} P.O. Box 3000, Boulder, Colorado 80307, USA}
 
 \date{\today}
 
\begin{abstract}
We compute solutions of the Lagrangian-Averaged Navier-Stokes
$\alpha-$model (\lans) for significantly higher Reynolds numbers (up
to $Re \approx 8300$) than have previously been accomplished.  This
allows sufficient separation of scales to observe a Navier-Stokes
inertial range followed by a second inertial range \neu{}{ specific to
\lans.  Both fully helical and non-helical flows are examined, up to
Reynolds numbers of $\sim 1300$}. The \neu{}{analysis of the}
third-order structure function scaling \neu{}{supports} the predicted
$l^3$ scaling; \neu{}{it corresponds} to a $k^{-1}$ scaling of the
energy spectrum for scales smaller than $\alpha$.  The energy spectrum
itself shows a different scaling which goes as $k^1$. This latter
spectrum is consistent with the absence of stretching in the
sub-filter scales due to the Taylor frozen-in hypothesis employed as a
closure in the derivation of \lans.  \neu{}{These two scalings are
conjectured to coexist in different spatial portions of the flow.  The
$l^3$ ($E(k)\sim k^{-1}$) scaling is subdominant to $k^1$ in the
energy spectrum, but} the $l^3$ scaling is responsible for the direct
energy cascade, as no cascade can result from motions with no internal
degrees of freedom.  We demonstrate verification of the prediction for
the size of the \lansa attractor resulting from this scaling. From
this, we give a methodology either for arriving at grid-independent
solutions for \lans, or for obtaining a formulation of \add{}{a} Large
Eddy Simulation (LES) \add{}{optimal in the context of the alpha
models}.  \neu{}{ The fully-converged grid-independent \lansa may not \resp{be}
the best approximation to a direct numerical simulation of the
Navier-Stokes equations since the minimum error is a balance between
truncation errors and the approximation error due to using \lansa
instead of the primitive equations. Furthermore, the small-scale
behavior of \lansa contributes to a reduction of flux at constant
energy, leading to a shallower energy spectrum for large
$\alpha$. These small-scale features, however, do not preclude \lansa
to reproduce correctly the intermittency properties of the high
Reynolds number flow.}

\end{abstract}
 
\maketitle

\section{Introduction}

Since the degrees of freedom for high Reynolds number ($Re$)
turbulence, as can be encountered in geophysical and
astrophysical flows, can be very large, the implementation of their
numerical modeling can easily exceed technological limits for
computations. Furthermore, since truncation of the omitted scales
removes important physics\add{}{, e.g., of multi-scale interactions, the only approach to a numerical study of such flows is to employ subgrid
modeling of those scales}.  This is frequently accomplished with Large
Eddy Simulations (LES--see \cite{M94,LM96,MK00} for recent reviews).
This is of importance for geophysical, astrophysical and engineering
applications and can have consequences for meteorological \cite{KSM99}
and climate prediction simulations \cite{H06}, for instance.  While
realistic Reynolds numbers will remain out of reach for the
foreseeable future, subgrid modeling can be an extremely useful tool
in the computation of simulations for such applications.

The incompressible Lagrangian-averaged Navier-Stokes equations (\lans,
$\alpha-$model, or also the viscous Camassa-Holm equation)
\cite{HMR98b,CFH+98,CFH+99a,CHM+99,CFH+99b,FHT01} is one possible
subgrid model.  It can be derived, for instance, by temporal averaging
applied to Hamilton's principle (where Taylor's frozen-in turbulence
hypothesis is applied as the closure, and also as the only
approximation of the derivation) \cite{HMR98a,H02a,H02b}.  For this
reason, the momentum-conservation structure of the equations are
retained.  For scales smaller than the filter width, \lansa reduces
the steepness of steep gradients of the Lagrangian mean velocity and
limits how thin vortex tubes become as they are transported (the
effect on larger length scales is negligible) \cite{CHM+99}.  The
$\alpha-$model may also be derived from smoothing the transport
velocity of a material loop in Kelvin's circulation theorem
\cite{FHT01}.  Consequently, there is no attenuation of resolved
circulation, which is important for many engineering and geophysical
flows where accurate prediction of circulation is highly desirable.
\lansa has previously been compared to direct numerical simulations
(DNS) of the Navier-Stokes equations at modest Taylor Reynolds numbers
($R_\lambda \approx 72$ \cite{ZM04}, $R_\lambda \approx 130$
\cite{CHM+99}, and $R_\lambda\approx 300$ \cite{CHO+05}).  \lansa was
compared to a dynamic eddy-viscosity LES in 3D isotropic turbulence
under two different forcing functions ($R_\lambda \approx 80$ and
$115$) and for decaying turbulence with initial conditions peaked at a
low wavenumber ($R_\lambda \approx 70$) and at a moderate wavenumber
($R_\lambda \approx 220$) \cite{MKS+03}.  \resp{In these comparisons,} \lansa was preferable in
that it demonstrated correct alignment between eigenvectors of the
subgrid stress tensor and the eigenvectors of the resolved stress
tensor and vorticity vector.  \lansa and a related regularization, the
Leray model, were contrasted with a dynamic mixed (similarity plus
eddy-viscosity) model in a turbulent mixing shear layer ($Re \approx
50$) \cite{GH02a,GH06}.  \lans, \resp{with relatively high subfilter
resolutions,} was the most accurate of these three LES \resp{tested at
this moderate $Re$}, but \resp{it was found that} the effects of numerical contamination can be
strong enough to lose most of this potential.  This could pose some
limitations on its practical use.  Quantifying those limitations is
one of the goals of this present work.  \resp{We will also find in this
study that, even with sufficient subfilter resolution, \lansa fails to
represent all the neglected physics in a more turbulent regime (higher
$Re$).}

The $\alpha-$model also describes an incompressible second-grade
non-Newtonian fluid (under a modified dissipation) \cite{FHT01}.  In
this interpretation, $\alpha$ is a material parameter which measures
the elastic response of the fluid.  Either from this standpoint, from
its status as a regularization of the Navier-Stokes equations, or,
independently of any physically motivation, as a set of partial
differential equations with proven unique regular solutions, we may
analyze \lansa without any LES considerations.  Analyzing
inertial-range scaling for \lansa for moderate and large $\alpha$, as
well as identifying different scalings at scales larger and smaller
than $\alpha$ is another of the goals of this work.  In this context
we also study the numerical resolution requirements to obtain
well-resolved solutions of \lansa (i.e., grid-independent solutions)
which leads to a verification of the predictions of the size of the
attractor in \lansa \cite{FHT01,GH06b}.  Section \ref{SEC:DETAILS}
presents the \lansa model, our numerical experiments and technique.
In Section \ref{SCALING} we analyze inertial-range scaling for \lans.
In Section \ref{GRIDINDEP} we determine the numerical resolution
requirements to obtain well-resolved solutions of \lans.  In Section
\ref{LES} we address the LES potential of \lansa by comparing
$\alpha-$model simulations to a $256^3$ DNS ($Re \approx 500$,
$R_\lambda\approx300$), a $512^3$ DNS ($Re \approx 670$,
$R_\lambda\approx350$), a $512^3$ DNS ($Re \approx 1300$,
$R_\lambda\approx490$), a $1024^3$ DNS ($Re \approx 3300$,
$R_\lambda\approx790$), and a $2048^3$ DNS ($Re \approx 8300$,
$R_\lambda\approx1300$). (The $Re \approx 3300$ simulation has been
previously described in a study of the imprint of large-scale flows on
local energy transfer \cite{AMP05b,MAP06}.)  In Section
\ref{RESULTS}, we compare and contrast in more detail \lansa solutions
with DNS at $Re\approx3300$.  Finally, in Section \ref{SEC:CONCL} we
summarize our results, present our conclusion, and propose future
directions of investigation.

\section{Technique}
\label{SEC:DETAILS}

We consider the incompressible Navier-Stokes equations for a fluid with
constant density,
\begin{eqnarray} \partial_tv_i +
v_j\partial_jv_i  = - \partial_ip + \nu \partial_{jj}v_i + F_i \nonumber
\\ \partial_iv_i = 0,
\thesis{\\ v_i = (1 - \alpha^2 \partial_{jj}) u_i,}
\label{eq:navier-stokes}
\end{eqnarray}
where $v_i$ denotes the component of the velocity field in the $x_i$
direction, $p$ the pressure divided by the density, $\nu$ the
kinematic viscosity, and $F_i$ an external force that drives the
turbulence (in all results, the time, $t$, is expressed in units of
the eddy-turnover time).  The \lansa equations
\cite{HMR98b,CFH+98,CFH+99a,CHM+99,CFH+99b,FHT01} are given by
\begin{eqnarray} \partial_tv_i +
u_j\partial_jv_i + v_j\partial_iu_j =-\partial_i\pi +  \nu
\partial_{jj}v_i + F_i \nonumber \\ \partial_iv_i =\partial_iu_i = 0,
\label{eq:lansSC}
\end{eqnarray}
where $u_i$ denotes the filtered component of the velocity field and
$\pi$ the modified pressure.  Filtering is accomplished by the
application of a normalized convolution filter $L: f \mapsto \bar{f}$
where $f$ is any scalar or vector field.  By convention, we define
$u_i \equiv \bar{v_i}$.  We choose as our filter the inverse of a
Helmholtz operator, $L = \mathcal H^{-1} = (1 -
\alpha^2\partial_{kk})^{-1}$.  Therefore, $\vec{u} = g_\alpha \otimes
\vec{v}$ where $g_\alpha$ is the Green's function for the Helmholtz
operator, $g_\alpha(r) = \exp (-r/\alpha)/(4\pi\alpha^2r)$ (i.e., the
well-known Yukawa potential), or in Fourier space, $\hat{\vec{u}}(k) =
\hat{\vec{v}}(k)/(1+\alpha^2k^2)$.

We solve Eqs. (\ref{eq:navier-stokes}) and (\ref{eq:lansSC}) using a
parallel pseudospectral code \cite{GMD05,GMD05b} in a
three-dimensional (3D) cube with periodic boundary conditions.
In most of the runs, we employ a Taylor-Green forcing
\cite{TG37},
\begin{equation}
F = \left[ \begin{array}{c}
    \sin k_0x \cos k_0y \cos k_0z \\
  - \cos k_0x \sin k_0y \cos k_0z \\
  0
  \end{array}\right]
\label{EQ:TGscale}
\end{equation}
(generally, with $k_0=2$), and employ dynamic control \cite{MPM+05} to
maintain a nearly constant energy with time.  This expression
Eq. (\ref{EQ:TGscale}) is not a solution of the Euler's equations, and as a
result small scales are generated fast when the fluid is stirred with
this forcing.
\thesis{That is, when substituted for $\vvec$, Eq. (\ref{EQ:TGscale}) does
not solve the Euler's equations.  Expanding the velocity response in time, the leading order will be proportional to the forcing and this motion rapidly excites smaller scale motions (see \cite{TG37}).}
The resulting flow models the fluid between
counter-rotating cylinders \cite{B90} and has been widely used to
study turbulence, including studies in the context of the generation
of magnetic fields through dynamo instability \cite{PMM+05}.  We also
consider some runs with random and ABC \cite{MAP06} forcing.  We
define the Taylor microscale as $ \lambda = 2\pi\sqrt{\langle
v^2\rangle/\langle \omega^2\rangle}, $ and the mean velocity
fluctuation as $ v_{rms} = \left( 2\int_0^\infty E(k) dk\right)^{1/2}.
$ The Taylor microscale Reynolds number is defined by $ R_\lambda =
{v_{rms}\lambda}/{\nu} $ and the Reynolds number based on a unit
length is $ Re = {v_{rms}\times1}/{\nu}$.

\section{Inertial range scaling of \lans}
\label{SCALING}

\subsection{$l^3$ scaling of third-order structure function derived from the K\'arm\'an-Howarth theorem for \lans}

For \lans, the $H^1_\alpha(u)$ norm is the quadratic invariant to be
identified with the energy,
\begin{equation}
\frac{dE_\alpha}{dt} =-2\nu\Omega_\alpha,
\label{EQ:LANSA_BALANCE}
\end{equation}
where
\begin{equation}
 E_\alpha = \frac{1}{D}\int_D\frac{1}{2}(\vec{u}-\alpha^2\nabla^2\vec{u})\cdot\vec{u} d^3x
= \frac{1}{D}\int_D\frac{1}{2}\vec{v}\cdot\vec{u} d^3x,
\end{equation}
and
\begin{equation}
\Omega_\alpha = \frac{1}{D}\int_D \frac{1}{2}\boldsymbol{\omega}\cdot\bar{\boldsymbol{\omega}} d^3x.
\end{equation}
As usual, we define the (omni-directional) spectral energy density,
$E_\alpha(k)$, from the relation
\begin{equation}
E_\alpha = \int_0^\infty \oint E_\alpha(\vec{k}) d\sigma d\vec{k} =
\int_0^\infty E_\alpha(k) dk
\end{equation}
where $\oint d\sigma$ represents integration over the surface of a
sphere.  The $\alpha-$model possesses a theorem corresponding to the
K\'arm\'an-Howarth theorem \cite{KH38} for the Navier-Stokes equations and, as in
the Navier-Stokes case, scaling of the inertial range energy spectra
may be derived from it \cite{H02c}.  We summarize here the dimensional
analysis argument for the \lansa inertial range scaling that follows
from this theorem, beginning from Equation (3.8) in Ref. \cite{H02c}.
We use the short notation $v_i \equiv v_i(\vec{x})$, $u_i^{'}
\equiv u_i^{'}(\vec{x}^{'},t)$ and $\vec{r} \equiv \vec{x}^{'} -
\vec{x}$.  In the statistically isotropic and homogeneous case,
without external forces and with $\nu=0$, taking the dot product of
Eq. (\ref{eq:lansSC}) with $u_j^{'}$ we can obtain the equation
\begin{equation}
\partial_t \paper{\mathcal Q_{ij}}\thesis{{{\mathcal Q}^\alpha}_{ij}} = \frac{\partial}{\partial_{r^m}}\left(
\paper{\mathcal T_{ij}^m}\thesis{{{\mathcal T}^\alpha}_{ij}^m} - \alpha^2 \paper{\mathcal S_{ij}^m}\thesis{{{\mathcal S}^\alpha}_{ij}^m} \right).
\label{eq:Holm38scale}
\end{equation}
The trace of this equation is the Fourier transform of the detailed
energy balance for \lans.
\begin{equation}
 \paper{\mathcal Q_{ij}}\thesis{{{\mathcal Q}^\alpha}_{ij}} = \left< v_iu_j^{'} + v_ju_i^{'}\right> 
\end{equation}
is the second-order correlation tensor while
\begin{equation}
\paper{\mathcal T_{ij}^m}\thesis{{{\mathcal T}^\alpha}_{ij}^m} = \left< (v_iu_j^{'} +v_ju_i^{'} +v_i^{'}u_j +v_j^{'}u_i )u^m\right>,
\end{equation}
and
\begin{equation}
\paper{\mathcal S_{ij}^m}\thesis{{{\mathcal S}^\alpha}_{ij}^m} = \left< (\partial_mu_l\partial_iu_l)u_j^{'}
+(\partial_mu_l\partial_ju_l)u_i^{'}
+(g_\alpha \otimes {\tau'}_j^m )v_i\\
+(g_\alpha \otimes {\tau'}_i^m )v_j\right>,
\end{equation}
are the third-order correlation tensors for \lansa and $\tau_i^j$ is
the sub-filter scale stress tensor.  For $\alpha=0$ this reduces to
the well-known relation derived by K\'arm\'an and Howarth.  The energy
dissipation rate for \lansa, $\varepsilon_\alpha$, satisfies
$\varepsilon_\alpha \propto \partial_t \paper{\mathcal Q_{ij}}\thesis{{{\mathcal Q}^\alpha}_{ij}}$.  By dimensional
analysis in Eq. (\ref{eq:Holm38scale}) we arrive at
\begin{equation}
\varepsilon_\alpha \sim \frac{1}{l}(vu^2 + \frac{\alpha^2}{l^2}u^3).
\label{EQ:FULL_SCALESC}
\end{equation}
For large scales such that $l \gg \alpha$, the second right hand
term is ignored, $\vec{u}\approx\vec{v}$,
$\varepsilon_\alpha\approx\varepsilon$, and we arrive at the scaling
of the four-fifths law, $<(\delta v_{\|}(l))^3> \sim \varepsilon l$
\cite{F95}.  Here, $\delta v_{\|}(l) \equiv
[\vec{v}(\vec{x+l})-\vec{v}(\vec{x})]\cdot\vec{l}/l$ is the
longitudinal increment of $\vec{v}$.  The four-fifths law expresses
that the third-order longitudinal structure function of $\vec{v}$,
${S}_3^v \equiv \langle(\delta v_{\|})^3\rangle$, is given in
the inertial range in terms of the mean energy dissipation per unit
mass $\varepsilon$ by
\begin{equation}
{S}_3^v = -\frac{4}{5}\varepsilon l,
\label{EQ:FOURFIFTHSlans}
\end{equation}
or, equivalently, that the flux of energy across scales in the inertial
 range is constant.  We also obtain the Kolmogorov 1941
 \cite{K41a,K41b,K41c} (hereafter, K41) energy spectrum, $E(k)k \sim
 v^2 \sim \varepsilon^{2/3}l^{2/3}$, or, equivalently,
\begin{equation}
E(k) \sim \varepsilon^{2/3}k^{-5/3}.
\end{equation}
For small scales such that $\thesis{\eta_K \ll }l \ll \alpha\thesis{ \ll l_F}$, however, \neu{}{$v \sim
\alpha^{2}l^{-2}u$ and both right hand terms are equivalent} in
Eq. (\ref{EQ:FULL_SCALESC}), and our scaling law becomes
\begin{equation}
{S}_3^u \equiv <(\delta u_{\|}(l))^3> \sim \varepsilon_\alpha\alpha^{-2} l^3.
\label{EQ:LCUBElans}
\end{equation}
\thesis{With sufficient scale separation, both scalings, Eqs. (\ref{EQ:FOURFIFTHSlans}) and (\ref{EQ:LCUBElans}), can be observed (see Fig. \ref{fig:Struct3}).}
\neu{}{Note that this scaling differs in a substantial way from the Kolmogorov scaling ($\sim l$).}
For our small scale energy spectrum we then have
\begin{equation}
E_\alpha(k)k \sim uv \sim \varepsilon_\alpha^{2/3}\alpha^{2/3},
\end{equation}
where we used $u \sim \alpha^{-2}l^{2}v$.  The energy spectrum for
scales smaller than $\alpha$ is then
\begin{equation}
E_\alpha(k)\sim \varepsilon_\alpha^{2/3}\alpha^{2/3}k^{-1}.
\label{EQ:LANSA_SPECTRUMSC}
\end{equation}
This spectrum can also be derived from phenomenological arguments
originally introduced by Kraichnan \cite{K67}, and it differs from the
Navier-Stokes spectrum due to the fact that the fluid is advected by
the smoothed velocity $\vec{u}$ which does not directly correspond to
the conserved energy $E_\alpha$ \cite{FHT01}.

\begin{figure}[htbp]\thesis{\begin{center}\leavevmode}
  \includegraphics[width=\paper{8.95}\thesis{12.5}cm]{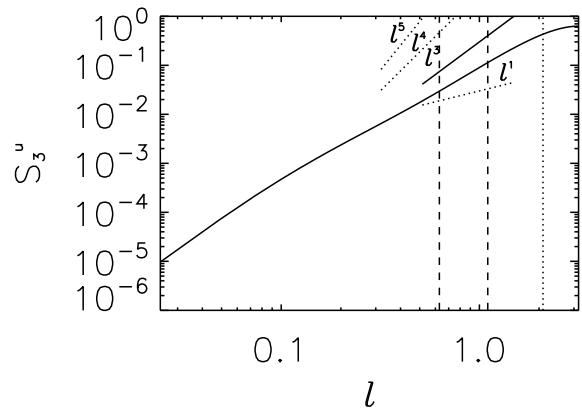}
  \caption[Third-order longitudinal structure function of the
  smoothed velocity field  versus length
  for large $\alpha$ \lans]{Third-order longitudinal structure function of the
  smoothed velocity field $\vec{u}$, ${S}_3^u$, versus $l$
  for large $\alpha$ \lansa ($\alpha=2\pi/3$ indicated by the vertical
  dotted line).  The scales identified with an inertial range are
  marked by vertical dashed lines and the scaling predicted by
  Eq. (\ref{EQ:LCUBElans}), $l^3$, is indicated by a solid line.  The
  fitted scaling exponent $\zeta_3^u$ (${S}_3^u(l)\sim
  l^{\zeta_3^u}$) is found to be $\zeta_3^u=2.39\pm.04$.  This is more
  consistent with the scaling given by Eq. (\ref{EQ:LCUBElans}) than K41
  scaling, $l^1$ Eq. (\ref{EQ:FOURFIFTHSlans}), or other proposed \lansa
  scalings (indicated by dotted lines, see text).}
  \label{FIG:KMINUS1}
\thesis{\end{center}}
\end{figure}

We test this prediction for \lansa scaling at a resolution of $256^3$
($\nu=1.2\times10^{-4}$) by moving both the forcing ($k_0=1$) and
$\alpha$ ($k_\alpha \equiv 2\pi/\alpha=3$) to large scales in order to
increase the number of resolved scales for which $k\alpha>1$.  In so
doing, we are assuming that the scaling for large $\alpha$ is the same
as for small $\alpha$ and large $k$ (for evidence to this effect, see
\cite{LKT+07}).  Confirmation as given by Eq. (\ref{EQ:LCUBElans}) is
presented in Fig. \ref{FIG:KMINUS1} where we plot ${S}_3^u$ as a
function of $l$ (by convention, we plot ${S}_3^u = <|\delta
u_{\|}(l)|^3>$ to reduce cancellation in the statistics).  The scales
identified with an inertial range $k\in[6,10]$ are marked by vertical
dashed lines and the predicted scaling, $l^3$, is indicated by a solid
line.  We fit a scaling exponent (${S}_3^u(l)\sim l^{\zeta_3^u}$) and
find $\zeta_3^u=2.39\pm.04$.  \neu{}{This is significantly steeper
than the classical Kolmogorov scaling given by
Eq. (\ref{EQ:FOURFIFTHSlans}); it can thus be viewed as more
consistent with the scaling given by Eq. (\ref{EQ:LCUBElans}). It is
also more consistent \neu{}{with $l^3$} than \neu{}{with} other possible \lansa scalings: under} the
assumption that the turnover time scale of eddies of size $\sim l$ is
determined by the unsmoothed velocity $\vec{v}$, we find
${S}_3^u(l)\sim l^5$, and if it is determined by
$\sqrt{\vec{v}\cdot\vec{u}}$, we find ${S}_3^u(l)\sim l^4$ (see, e.g.,
Refs. \cite{LKT+07,CHT05,CHO+05,ILT06}).  The observed scaling
corresponds to none of these cases, and is actually \neu{}{closer to
an evaluation of the} turnover time $t_l$ at the scale $l$ given by
$t_l \sim l / u_l$ (with ${S}_3^u(l)\sim l^3$).  Note that for 2D
\lans, however, it is the case that the scaling is determined by the
unsmoothed velocity $\vec{v}$ \cite{LKT+07}.  We note that this is one
of many differences between the 2D and 3D cases (e.g., ideal
invariants and cascades).  Another difference, which we shall show in
Section \ref{RESULTS}, is that in 2D vorticity structures decrease in
scale as $\alpha$ increases while in 3D there is a change in aspect
ratio with structures getting both shorter and fatter.  This may, in
fact, be related to the shallower \lansa energy spectrum for
$k\alpha>1$ which we show in Section \ref{RESULTS}.  While
differences are observed between the scaling shown in
Fig. \ref{FIG:KMINUS1} and Eq. (\ref{EQ:LCUBElans}), the error bars
\neu{}{deny} a K41 scaling (as well as the $l^4$ and $l^5$ scalings) at
scales smaller than $\alpha$.  We believe \neu{}{the discrepancy
between the observed and predicted scaling} can be due to
lack of resolution to resolve properly the inertial range at
sub-filter scales.  We have less than a decade of inertial range and
only $256^3$ points for the statistics.  As more computational
resources become available, this scaling should be re-examined.

\subsection{Subdominance of the $k^{-1}$ energy spectrum and rigid-body motions}
\label{SUBDOMINANT}

\begin{figure}[htbp]\thesis{\begin{center}\leavevmode}
  \includegraphics[width=\paper{8.95}\thesis{12.5}cm]{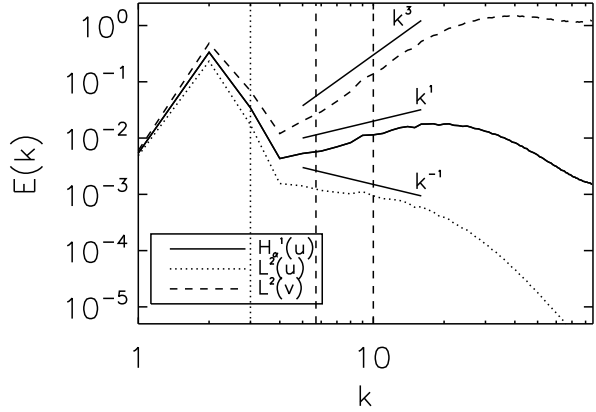}
  \caption[Spectral energy density versus wavenumber for
    large$-\alpha$ \lansa] {Spectral energy density, $E(k)$, versus
    wavenumber, $k$, for large$-\alpha$ \lansa solution.  Here forcing
    ($k_0=1$) and $\alpha$ ($k_\alpha\equiv2\pi/\alpha=3$, vertical dotted line) are
    set at the largest scales to increase the number of scales for
    which $k\alpha>1$.  Spectra are plotted for three norms:
    $H^1_\alpha(u)$ norm (solid line), $L^2(u)$ norm (dotted line),
    and the $L^2(v)$ norm (dashed line).  As these last two norms are
    not quadratic invariants of \lans, we employ the $H^1_\alpha$ norm
    for all following results.  All three spectra correspond to that
    derived from the assumption of rigid {bodies} in the
    smoothed velocity $\vec{u}$, Eq. (\ref{EQ:RIGID_SPECTRUM}).  The
    vertical dashed lines are at the same scales as those in
    Fig. \ref{FIG:KMINUS1}.}
  \label{FIG:NORMS}
\thesis{\end{center}}
\end{figure}

As a consequence of \lans's Taylor's frozen-in hypothesis closure,
scales smaller than $\alpha$ can phase-lock into coherent structures
and be swept along by the larger scales (see, e.g., \cite{H02c}).  If
we assume, formally, that this ``frozen-in turbulence'' takes the form
of ``rigid \neu{}{bodies}'' in the smoothed velocity field (no
stretching), we arrive at a much different spectrum than $k^{-1}$,
Eq. (\ref{EQ:LANSA_SPECTRUMSC}).
\neu{}{All scales smaller than $\alpha$ are subject to the frozen-in
hypothesis and we expect to find such rigid bodies at these scales.}
We note that \neu{}{collections of
``rigid'' portions of the flow (rotating or non-rotating)} reduce the
total degrees of freedom (\dof) and make physical sense with \lans's
relation to second-grade fluids: these rigid \neu{}{bodies} can be
envisioned as polymerized portions of the fluid.  As a matter of fact,
in such structures all internal {\sl dof} are frozen.  \neu{}{These ``rigid bodies'' follow as well from the consideration of \lansa as an initial value
problem in Fourier space, for which we have $\hat{\vec{u}}(k) =
\hat{\vec{v}}(k)/(1+\alpha^2k^2)$.  In the limit as $\alpha$ approaches
infinity, all wavenumber (and spatial) dependence for $\uvec$ is eliminated
and the entire flow is advected by a uniform velocity field (advection without internal degrees of freedom).}

For a rigid
\neu{}{body} there can be no stretching and, therefore, all the
longitudinal velocity increments, $\delta u_{\|}$, must be identically
zero ($\vec{\delta u}(\vec{l}) = \boldsymbol{\Omega}\times\vec{l}$ from basic
mechanics with $\boldsymbol{\Omega}$ the rotation vector and, hence, $\delta
u_{\|}(l) = \vec{\delta u}(\vec{l})\cdot\vec{l}/l = 0$).  Note that in
\lansa Eq. (\ref{eq:lansSC}) the $v_j\partial_iu_j$ term contributes
only a rotation and not a stretching of $\vec{u}$.  Such
polymerization would have two consequences.  Firstly, since there is
no stretching, these rigid \neu{}{bodies} would not contribute to the
turbulent energy cascade,
\begin{equation}
<(\delta u_{\|}(l))^3> = 0.
\end{equation}
Secondly, the energy spectrum from dimensional analysis ($u^2 \sim
\mbox{const}$, for large $\alpha/l$: $u=(1+\alpha^2/l^2)^{-1}v \sim
l^2v$, and $E_\alpha(k)k \sim uv \sim k^2$) is
\begin{equation}
E_\alpha(k) \sim k.
\label{EQ:RIGID_SPECTRUM}
\end{equation}
This is, in fact, the observed \lansa spectrum for $k\alpha\gg1$ as is
shown in Fig. \ref{FIG:NORMS}.  We verified that the spectrum is not
the result of under-resolved runs, as is the case, e.g., in the $k^2$
spectrum observed in truncated Euler systems \cite{CBD+05} or in
extremely under-resolved spectral simulations of the Navier-Stokes
equations.  Indeed, equipartition of the energy among all modes in a
truncated Euler$-\alpha$ system should also lead to a $k^2$ spectrum.
Along with several experiments with different viscosities and also
with statistically homogeneous and isotropic forcing (not shown here),
these are assurances that the observed spectrum is not a result of
inadequate numerical resolution.  It should be noted that this is the
same computation for which the third-order structure function is shown
in Fig. \ref{FIG:KMINUS1}.  The third-order structure function is
consistent with a $l^3$ scaling (corresponding to a $k^{-1}$ energy
spectrum) while the spectrum itself is $k^{1}$.  (Also shown in
Fig. \ref{FIG:NORMS} are the $L^2(u) \equiv \langle u^2\rangle/2$ and
the $L^2(v) \equiv \langle v^2\rangle/2$ norms which (through $u\sim
\alpha^2v/k^2$ for $k\alpha\gg1$) correspond to $k^{-1}$ and $k^{3}$
spectra, respectively.  Since the analytical properties of the \lansa
solution are based on the energy balance, ${dE_\alpha}/{dt}
=-2\nu\Omega_\alpha$, in the $H^1_\alpha(u)$ norm, we employ this
norm for all following results.)  These two different scalings, $l^3$
and $k^{1}$, are consistent with a picture where a fluid has both
\neu{}{rigid-body portions at scales smaller than $\alpha$} (wherein there is no turbulent cascade) and \neu{}{spatial}
regions between these where the cascade does take place.  For the
structure functions, a non-cascading rigid \neu{}{body} does not contribute
to the scaling and consequently the cascading contribution,
Eq. (\ref{EQ:LCUBElans}), dominates.  The energy spectrum, however,
\neu{}{for the limit of $k$ very large,}
 is
dominated by the $k^{+1}$ term, and hence
the $k^{-1}$ component is subdominant.

\begin{figure}[htbp]
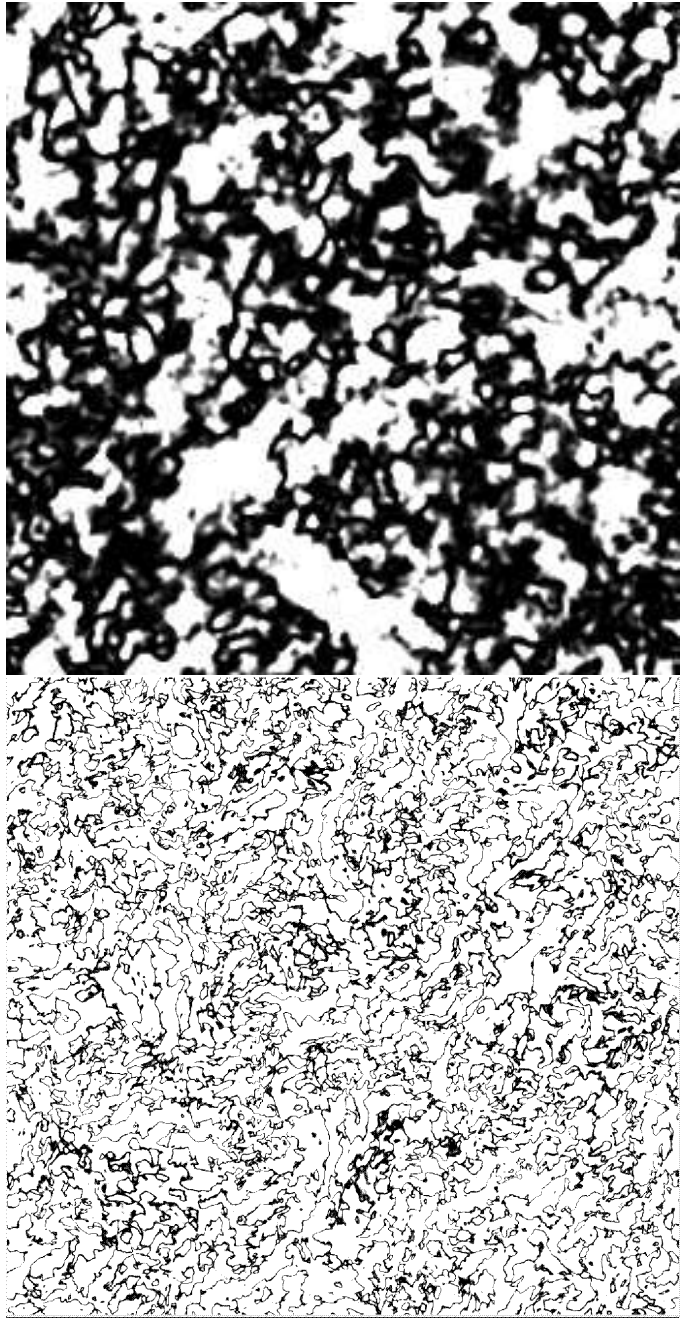

\thesis{\begin{center}\leavevmode}
  \includegraphics[width=8.95cm]{\thesis{scaling/}fig3a}\\
  \includegraphics[width=8.95cm]{\thesis{scaling/}fig3b}
  \caption[Two-dimensional slice of the cubed longitudinal
  increment] {Two-dimensional slice of the cubed longitudinal
  increment $(\delta u_{\|}(2\pi/10))^3$for \lansa and $(\delta
  v_{\|}(2\pi/10))^3$ for DNS.  For all black pixels, the cubed
  longitudinal increment is less than $10^{-2}$ (approximately
  consistent with rigid {bodies}).  On the top is the large-$\alpha$
  simulation ($k_0=1$, $k_\alpha=3$, $\nu=1.2\times10^{-4}$) where the
  filling factor (computed over the entire 3D domain) is 0.67.  On the
  bottom is a DNS of Navier-Stokes ($k_0=2$, $\nu=3\times10^{-4}$) where
  the filling factor is 0.26.  Thus, a much greater portion of the
  flow is consistent with {collections of rigid bodies} for the large$-\alpha$
  simulation.}
  \label{FIG:RIGID_ROTATE}
\thesis{\end{center}}
\end{figure}

\begin{figure}[htbp]\thesis{\begin{center}\leavevmode}
  \includegraphics[width=\thesis{12.5}\paper{8.95}cm]{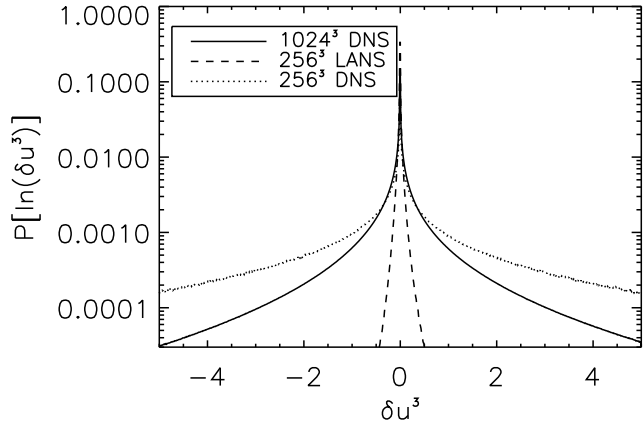}
  \caption[Pdfs of cubed longitudinal increment] {Pdfs of $(\delta v_{\|}(2\pi/10))^3$ for DNS ($N=1024$,
solid line), and of $(\delta u_{\|}(2\pi/10))^3$ for \lansa ($N=256$,
dashed line), and of the DNS downgraded to lower resolution ($N=256$,
dotted line).  See Fig. \ref{FIG:RIGID_ROTATE} for simulation
parameters.  Note that both {\pdf}s have a slight positive asymmetry
consistent with a positive dissipation rate $\varepsilon_{(\alpha)}$.
The \lansa \pdfa is more strongly concentrated around zero consistent
with the idea that portions of the flow (at scales smaller than
$\alpha$) are {acting as rigid bodies}.}
  \label{FIG:RIGID_PDFS}
\thesis{\end{center}}
\end{figure}

We further explore the validity of this picture by examining the
spatial variation of the cubed longitudinal increment, $(\delta
v_{\|}(l))^3$ in DNS, and $(\delta u_{\|}(l))^3$ in \lansa for
$\alpha/l \gg 1$, which in each case is proportional to the energy
flux across a fixed scale $l$.
\neu{}{(The presence of the hypothesized ``rigid bodies'' should be evident
as significant portions of the flow where there is no energy flux.)}
  In Fig.  \ref{FIG:RIGID_ROTATE} we
show visualizations of these quantities corresponding to $l=2\pi/10$
($k=10$) for both the large-$\alpha$ \lansa simulation and a highly
turbulent DNS ($k_0=2$, $\nu=3\times10^{-4}$).  The scale ($k=10$) is
chosen as it is in the inertial ranges of both flows.  We note that
for \lans, a significant portion of the flow is not contributing to
the flux of energy to smaller scales (the filling factor for $(\delta
u_{\|}(2\pi/10))^3 < 10^{-2}$ is 0.67 as compared to 0.26 for the
Navier-Stokes case).  These regions can be identified as
``polymerized'' or ``rigid \neu{}{bodies}'' in $\vec{u}$ \neu{}{and their locations are found to be robust when the $l$ used for $(\delta u_{\|}(l))^3$ is varied over a factor of 2}.  Moreover, this
is highlighted in the probability distribution functions ({\pdf}s), see
Fig. \ref{FIG:RIGID_PDFS}, where we see the \lansa \pdfa is more strongly
concentrated around zero than the DNS.  This is consistent with the
idea that the internal {\sl dof} of large portions of the flow (at
scales smaller than $\alpha$) are frozen.  We point out that this
comparison is not a LES validation, but, rather, a comparison between
the dynamics of two different fluids at similar Reynolds numbers.  One
flow is a well-resolved numerical solution of the Navier-Stokes
equations, and the other is a well-resolved solution of the \lansa
equations with large $\alpha$.  For this reason a reduced resolution
($N=256$) representation for the DNS (for which $N=1024$) is not
depicted in Fig.  \ref{FIG:RIGID_ROTATE}.  We have performed such a
down-sampling, however, and find the filling factor is reduced even
more, to 0.14, and the tails of the \pdfa increase over the
full-resolution analysis (dotted line in Fig. \ref{FIG:RIGID_PDFS}).
No inverse Helmholtz filtering, $\mathcal H^{-1}$ is applied to the
DNS data.  Note that this would amount to computing $(\delta
u_{\|}(l))^3$ in the DNS, which has no meaning in the dynamics of the
Navier-Stokes equations (the energy flux is proportional to $(\delta
v_{\|}(l))^3$).

\begin{figure}[htbp]\thesis{\begin{center}\leavevmode}
  \includegraphics[width=\thesis{12.5}\paper{8.95}cm]{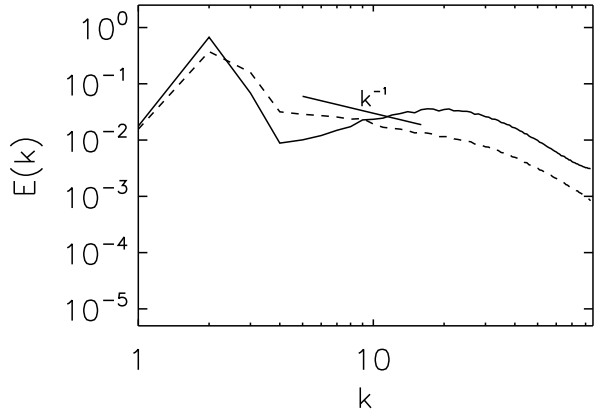}
  \caption[Coexistent energy spectra: $k^{-1}$ and $k^1$]
{{Spectral energy density, $E(k)$, versus wavenumber, $k$, for
large$-\alpha$ \lansa solution.  The solid line indicates the spectrum
as given in Fig. \ref{FIG:NORMS} but for a single snapshot (the same
as selected for Fig. \ref{FIG:RIGID_ROTATE}).  The dashed line
indicates the spectrum wherein all portions of the flow associated
with ``rigid bodies'' (a 2D slice of which is shown in
Fig. \ref{FIG:RIGID_ROTATE}) are removed.  This provides further
evidence that the flow spatially in between the ``rigid bodies''
possesses a negative power law energy spectrum (the predicted $k^{-1}$
power law is shown as a solid line).}}
  \label{FIG:TWO_SPECTRA}
\thesis{\end{center}}
\end{figure}

\neu{}{We end this section with further evidence of coexistent energy
spectra, $k^{-1}$ and $k^1$, in separate spatial portions of the flow.
We mask out all portions of the flow that we identify with rigid
bodies ($(\delta u_\|(2\pi/10))^3 < 10^{-2}$, a 2D slice of which is
shown in Fig. \ref{FIG:RIGID_ROTATE}).  The energy spectrum of the
remaining portion of the flow is shown in Fig. \ref{FIG:TWO_SPECTRA}
as a dashed line to be compared with the spectrum of the entire flow
shown as a solid line.  The operation of spatially filtering the flow
before computing the spectrum serves to ``smear out'' the energy
spectrum by convolving it with the spectrum of the filter.
Deconvolution in 3D with $N=256$ is intractable and we are, therefore,
unable to remove this ``smearing'' of the energy spectrum of the
cascading portions of the flow.  Nonetheless, after conducting what
tests we could with the filtering process (not shown here), we
conclude that the power law of the energy spectrum of these portions
is negative and, thus, distinctly different from that of the rigid bodies.}

\section{Resolution requirements for grid-independent \lansa solutions: Size of attractor}
\label{GRIDINDEP}

It is useful to make a distinction between the quality of a subgrid
model and effects arising from nonlinear interactions with
discretization errors at marginal spatial resolutions (which are more
characteristic of the discretization employed than of the subgrid
model) \cite{GF02,MGB03,GH06}.  Before doing this, we require an
estimate for the total degrees of freedom for the \lansa attractor
which as we show, unlike for the 2D case (see \cite{LKT+07}), for
the 3D case is reduced compared to Navier-Stokes.  The subdominant
$l^3$ scaling is associated with the flux of energy to small scales
and thus must be used to estimate the degrees of freedom of the \lansa
attractor, $\mbox{\sl dof}_\alpha$.  For dissipation the large
wavenumbers dominate and, therefore, combining the \lansa energy
balance, Eq. (\ref{EQ:LANSA_BALANCE}), with its sub-filter scale energy
spectrum, Eq. (\ref{EQ:LANSA_SPECTRUMSC}), allows us to implicitly specify
its dissipation wavenumber, $k_\eta^\alpha$, by
\begin{equation}
\frac{\varepsilon_\alpha}{\nu} \sim \int^{k_\eta^\alpha} k^2 E_\alpha(k)
 dk \sim \int^{k_\eta^\alpha} k^2
 {\varepsilon_\alpha}^{2/3}\alpha^{2/3}k^{-1} dk\\ \sim
{\varepsilon_\alpha}^{2/3}\alpha^{2/3}({k_\eta^\alpha})^2.
\end{equation}
Then we have,
\begin{equation}
k_\eta^\alpha \sim \frac{{\varepsilon_\alpha}^{1/6}}{\nu^{1/2}\alpha^{1/3}}.
\end{equation}
Using that the linear numerical resolution, $N$, must be proportional
to the dissipation wavenumber ($N \geq 3k_\eta^\alpha$) and that $Re
\sim \nu^{-1}$, we arrive at
\begin{equation}
N =C_0 {k_\alpha}^{1/3}Re^{1/2},
\label{EQ:THEORY}
\end{equation}
or, equivalently,
\begin{equation}
\mbox{\sl dof}_\alpha =\frac{C_0^3}{27\alpha}Re^{3/2},
\end{equation}
where $C_0$ is an unknown constant (for further details see
\cite{FHT01}).  We verify this prediction and determine the constant
$C_0$ through the use of a database stemming from studies in which
both the free parameter, $\alpha$ (or, equivalently, $k_\alpha$) and
the linear resolution, $N$, for a set of DNS flows with $Re \approx
500$, $670$, $1300$, and $3300$ are varied.  In so doing, we establish
the necessary numerical resolution for convergence to a
grid-independent solution.

\begin{figure}[htbp]
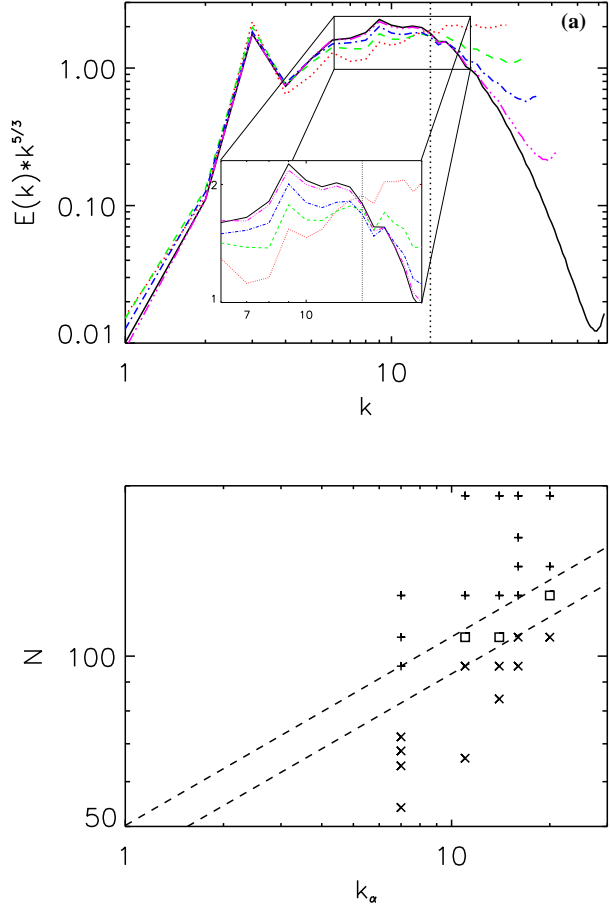

\thesis{\begin{center}\leavevmode}
  \includegraphics[width=\paper{8.95}\thesis{11}cm]{\thesis{scaling/}fig6a}
  \includegraphics[width=\paper{8.95}\thesis{11}cm]{\thesis{scaling/}fig6b}
  \caption[Plots for $Re\approx500$ simulations
demonstrating convergence to the grid-independent \lansa solution] {\paper{(Color online.)}  Plots for $Re\approx500$ simulations
demonstrating convergence to the grid-independent \lansa solution.
{\bf(a)} Average energy spectra ($t\in [20,33]$, $t$ is time in units
of eddy turn-over time) compensated by K41 for \lansa simulations,
$k_\alpha = 14$: $192^3$ (black solid), $84^3$ (red
dotted), $96^3$ (green dashed), $108^3$ (blue dash-dotted), and
$128^3$ (pink dash-triple-dot).  The vertical dashed line denotes
$k_\alpha$.  Inset is a blow-up near $k_\alpha$ where convergence can
be clearly seen.  \lansa at a linear resolution of $128^3$ is
approximately converged to the grid-independent solution while
resolutions of $96^3$ and less are clearly not.  {\bf(b)} The linear
resolution of $\alpha-$model simulations, $N$, is plotted versus
$k_\alpha$.  Simulations with inadequate resolution are plotted as
X's, those with approximately grid-independent solutions as +'s, and
experiments that are neither clearly resolved nor clearly unresolved
as boxes.  The dashed lines represent {$N=Ck_\alpha^{1/3}$} indicating
that a {constant} in the range $43.2 < C < 50.2$ agrees
with our data.  This partially confirms the prediction of
Eq. (\ref{EQ:THEORY}) and provides a reliable method to determine the
needed resolution for a grid-independent \lansa solution at a fixed
$Re$.}
  \label{FIG:SCALE_TEST_256}
\thesis{\end{center}}
\end{figure}

\begin{figure}[htbp]\thesis{\begin{center}\leavevmode}
  \includegraphics[width=\paper{8.95}\thesis{12.5}cm]{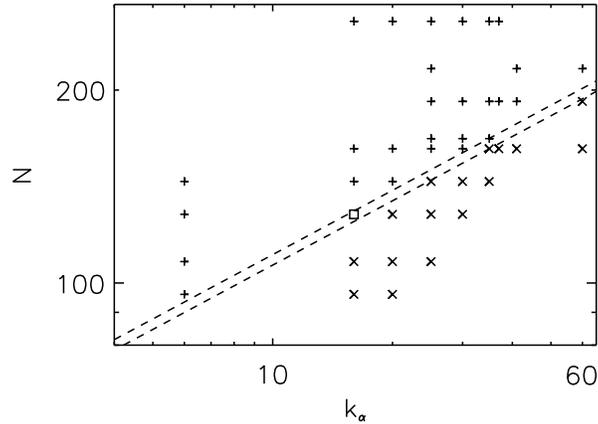}
  \caption[$Re\approx670$ simulations: linear
resolution of $\alpha-$model simulations versus $\alpha^{-1}$, convergence results] {As Fig. \ref{FIG:SCALE_TEST_256}(b) but for
$Re\approx670$ simulations.  The dashed lines represent
{$N=Ck_\alpha^{1/3}$} indicating that a {constant} in the
range $49.5 < C < 51.4$ agrees with our data.  {Note also that any power law, $N\propto k_\alpha^\beta$, with $0.30 < \beta < 0.46$ also agrees with the data.}}
  \label{FIG:SCALE_TEST_512}
\thesis{\end{center}}
\end{figure}

Convergence to the grid-independent solution is determined by
comparison of the energy spectrum, $E_\alpha(k)$, between runs with a
constant filter and varying resolution.  In
Fig. \ref{FIG:SCALE_TEST_256}(a), we make such a comparison for
$Re\approx500$ ($N=256$ for DNS) and $k_\alpha=14$ ($N=84$, $96$,
$108$, $128$, and $192$ for \lans).  We plot energy spectra
compensated by $k^{5/3}$ so that a K41 $k^{-5/3}$ spectrum would be
flat.  We see, based on comparing the energy spectra at wavenumbers
smaller than $k_\alpha$ to the $192^3$ \lansa spectrum, that
simulations at resolutions of $96^3$ and less are not converged while
the one at $128^3$ is.  That is, except for the very small scales at
the end of the dissipative range, there is very little difference
between the spectra at $128^3$ and at $192^3$ (i.e., the solution is
``grid-independent'').  Meanwhile, for resolutions of $96^3$ and less
the spectra vary greatly with resolution (i.e., they are
``unresolved'').  In Fig. \ref{FIG:SCALE_TEST_256}(b), we collect all
the results of similar studies ($Re\approx500$) in a plot of
resolution, $N$, versus inverse filter width, $k_\alpha$.  (We change
$N$ for a given $\alpha$, then change $\alpha$ and iterate.)  Pluses
correspond to grid-independent solutions, X's to under-resolved
solutions, and squares to ``undecided'' runs (i.e., that are neither
clearly resolved nor clearly under-resolved).  The dashed lines
represent Eq. (\ref{EQ:THEORY}) with the minimal and maximal choice of
$C$ (where $C_0 = CRe^{1/2}$), that agrees with our results (i.e.,
$43.2<C<50.2$).  In Fig. \ref{FIG:SCALE_TEST_512} we conduct similar
studies for $Re\approx670$.  We find $49.5<C<51.4$ and again validate
the predictive power of Eq. (\ref{EQ:THEORY}) for the necessary
numerical resolution for grid-independent solutions.

\begin{figure}[htbp]\thesis{\begin{center}\leavevmode}
  \includegraphics[width=\paper{8.95}\thesis{12.5}cm]{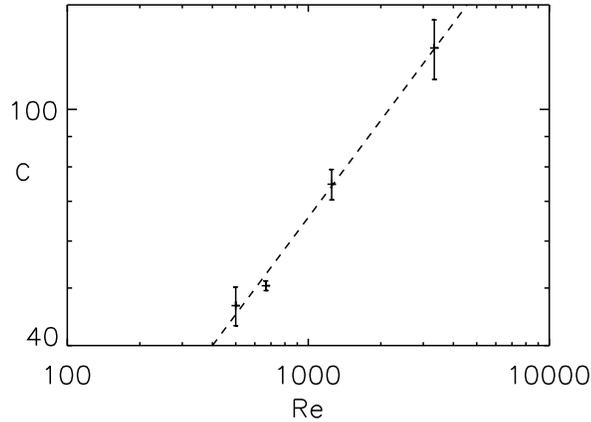}
  \caption[Scaling with $Re$ of \dofa for \lans] {Acceptable choices of {$C = C_0 Re^{1/2}$,} versus Reynolds number, $Re$,
  for grid-independent \lans.  {Error bars are not confidence levels, but} depict the range of values
  consistent with our database {($N=Ck_\alpha^{1/3}$)} at the four Reynolds numbers we tested.
  The dashed line depicts the least-squares fit with slope
  $0.54\pm0.14$.  This completes the validation of
  Eq. (\ref{EQ:THEORY}) which predicts $0.5$.  }
  \label{FIG:SCALE_CONFIRM}
\thesis{\end{center}}
\end{figure}

The greatest utility of the prediction, however, is due to the single
constant $C_0$ which is independent of Reynolds number.  A
determination of this constant can cheaply be achieved repeating this
process for several runs for low and moderate $Re$, and determines the
resolution requirement for the highest $Re$ attainable.  The ranges of
acceptable constants, $C = C_0 Re^{1/2}$, for the four Reynolds number
flows studied are plotted versus $Re$ in Fig. \ref{FIG:SCALE_CONFIRM}.
A power law $C = C_0 Re^{\gamma}$ fits our data with $\gamma =
0.54\pm0.14$ demonstrating the final validation of the prediction,
$\gamma = 0.5$, Eq. (\ref{EQ:THEORY}).  The value of the constant is
found to be $C_0 = 2.0 \pm 0.2$.  We made one study for the
maximally-helical ABC forcing at $Re \approx 1600$ and $\alpha =
2\pi/25$.  It is consistent with a value of $C_0 = 1.8 \pm 0.1$.  We
therefore conclude that the constant $C_0$ is not a strong function of
the forcing employed or of the scale at which the system is forced.
As a result, and unlike in 2D \lansa \cite{LKT+07}, we verify that the
size of the attractor in 3D \lansa is smaller than that in Navier-Stokes,
which is a promising result if the \lansa equation is going to be used
as an LES.  However, before doing this,  an assessment of the
truncation errors introduced in discretized systems (as used to
solve the equations numerically) and a study of the optimal choice for
$\alpha$ to capture the properties of a DNS is needed.  We consider
these problems in the following section.

\section{Can \lansa be considered as a Large Eddy Simulation?}
\label{LES}

In this section, we consider the \lansa equations as a means to an end,
and consider the solutions to their discretized equations as
approximations to the Navier-Stokes solutions.  We seek numerical
approximations of \lansa that minimize the difference to a fully
resolved or direct numerical solution (DNS) of Navier-Stokes \add{}{(i.e., we
analyze the behavior of  \lansa solutions in the LES framework, and call here the model a ``\lansa LES'', or in short ``$\alpha$-LES'').
In the LES framework, \lans's turbulent stress tensor, ${\bar\tau^\alpha}_{ij}$, is given by
(see, e.g., \cite{HN03})
\begin{eqnarray}
{\bar\tau^\alpha}_{ij} = \mathcal H^{-1} \alpha^2 (\partial_k{u}_i\partial_k{u}_j +
\partial_k{u}_i\partial_j{u}_k - \partial_i{u}_k\partial_j{u}_k).
\end{eqnarray}
}Previous studies have not made the
distinction between grid-independent \lansa and \lansa LES, though one
did study convergence to grid-independent solutions at moderate $Re$
\cite{GH06}.  We find, however, a definite difference between the two
approaches.  We show in this section that, in fact, \lansa
combined with truncation error yields a better fit to DNS than
grid-independent \lans.  The resolution that yields an \add{}{optimal $\alpha$-LES}
\add{}{(a terminology to be defined below)}
is also found to follow Eq. (\ref{EQ:THEORY}).  In the Section
\ref{LES_QUALITY}, we then address 
\add{}{the quality and usability of the predictions of the \lansa model viewed as an LES.}

\add{}{A remark about nomenclature may be in order at this point.
Traditionally, and for good reasons, LES attempt at capturing the large-scale properties of a flow with a huge Reynolds number, as found, e.g., in the atmosphere. In that case, the wavenumber at which the DNS is truncated is, at best, in the inertial range and it might even be in the energy-containing range, as 
for  the atmospheric boundary layer with a Taylor Reynolds number $R_{\lambda}\sim 10^4$. Of a different nature are the modeling methods sometimes called quasi-DNS. Here, the idea is to model a flow at a given, moderate Reynolds number but with an expense in computing resources lesser than if performing a DNS. Under-resolved DNS fall in that category; in that case, the large-scales are presumably well reproduced but the small scales are noisy. It is in that spirit that we now examine the properties of the \lansa model.}
\add{}{We thus qualify a model as optimal in the sense of being optimal for the class of 
\lansa models examined herein; in order to avoid repetition, we also use the terminology of alpha-optimal.
}

\begin{figure}[htbp]\thesis{\begin{center}\leavevmode}
  \includegraphics[width=\paper{8.95}\thesis{12.5}cm]{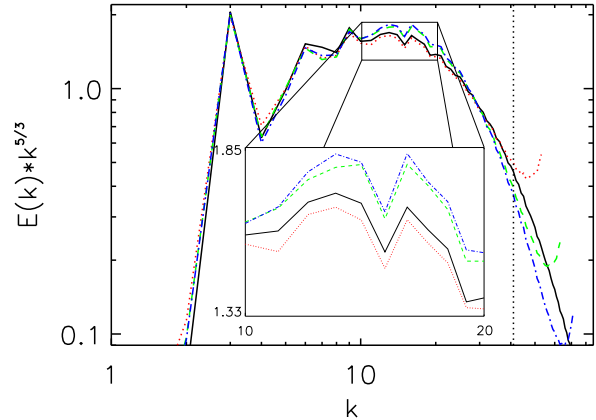}
  \caption[Plot of $Re\approx670$ simulations demonstrating optimal
$\alpha-$LES] {\paper{(Color online.)}  Plot of $Re\approx670$ simulations. Average
compensated energy spectra: DNS (solid black line) and \lansa
simulations, $k_\alpha = 41$: $N= 162$ (red dotted), $N=192$ (green
dashed), and $N= 216$ (blue dash-dotted).  \lansa at a linear
resolution of 192 is approximately converged to the grid-independent
solution while a resolution of 162 is not. $N=162$ does correspond,
however, more closely to the DNS spectrum.  We observe, in general,
that a combination of \lansa and truncation error yields the optimal {$\alpha$-}LES.}
  \label{FIG:512_LES}
\thesis{\end{center}}
\end{figure}

\begin{figure}[htbp]
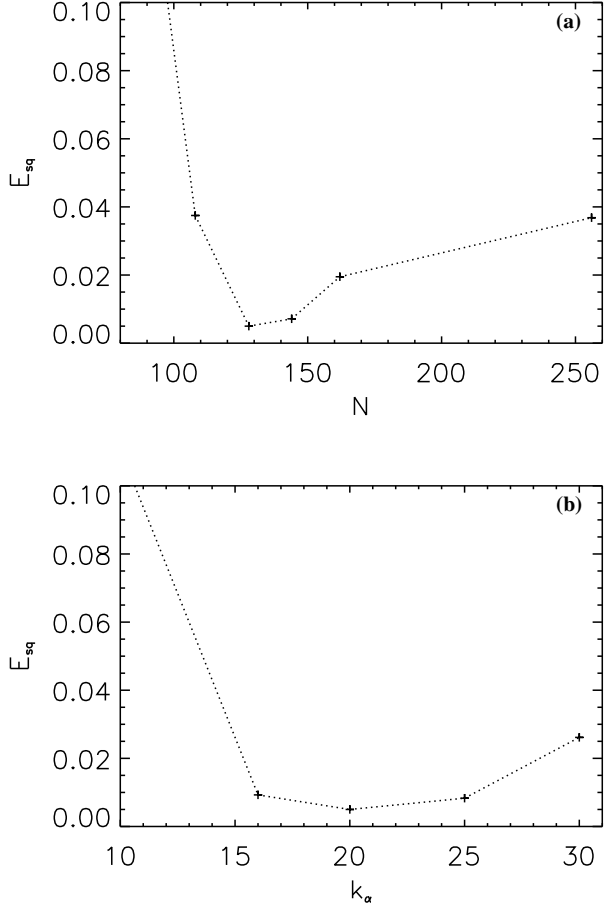
\thesis{\begin{center}\leavevmode}
  \includegraphics[width=\paper{8.95}\thesis{12.5}cm]{\thesis{scaling/}fig10a}
  \includegraphics[width=\paper{8.95}\thesis{12.5}cm]{\thesis{scaling/}fig10b}
  \caption[Plots for $Re\approx670$ simulations: Error versus
simulation resolution and versus increasing filter width] {Plots for $Re\approx670$
simulations.  {\bf(a)} Error (see Eq. (\ref{EQ:ERR_2SC})) versus
simulation resolution for $k_\alpha=20$.  The optimal (grid-dependent)
LES is for a resolution of $N\approx128$ and has a much smaller error
compared to the DNS than the grid-independent \lansa solution at
higher resolution.  {\bf(b)} Error versus $k_\alpha$ for
$N=128$.  At a given resolution the optimal value for $\alpha$ is not
zero but occurs at a local minimal error.  Any $k_\alpha\in[15,25]$
has an error near the minimum indicating that an LES solution may
perform well for a range of parameters near the optimal ones.  A
constant of $C = C_0 Re^{1/2} \approx 47$ in Eq. (\ref{EQ:THEORY}) is
found to correspond with optimal {$\alpha$-}LES approximations.}
  \label{FIG:LES_ERR}
\thesis{\end{center}}
\end{figure}

In Fig. \ref{FIG:512_LES} with $k_\alpha=41$, we plot the
$Re\approx670$ DNS spectrum (solid black line) and \lansa spectra at
three different resolutions.  We observe that, while the $N=162$
solution \paper{(dotted line, red online)}\thesis{(dotted red line)} is not converged, it is a better
approximation to the DNS than the grid-independent \lansa solution.
For all simulations we studied, \neu{}{the grid-independent \lansa solution is not the best approximation to the DNS}.  Another
example is given in Fig. \ref{FIG:LES_ERR} where we plot the
mean square spectral error normalized to make fair comparisons between
large and small $k_\alpha$ results,
\begin{equation}
E_{sq} = \frac{1}{n}\sum_{k=k_F}^{k_\alpha}\frac{(E_{\alpha}(k)-E(k))^2}
{E^2(k)},
\label{EQ:ERR_2SC}
\end{equation}
where $k_F$ is the wavenumber for the forcing scale, $E(k)$ is the DNS
spectrum (in the $L^2(v)$ norm), $E_\alpha(k)$ is the \lansa spectrum
(in the $H^1_\alpha(u)$ norm), and $n$ is the number of terms in the
sum.  These errors are calculated for spectra averaged over turbulent
steady-state solutions: $t \in [16,19]$ for $Re \approx 670$.  We see
that for a given filter or a given simulation resolution, there is a
local minimum in the error.  This minimum is a balance between
truncation errors and the approximation error due to using \lansa
instead of the full Navier-Stokes equations.  Due to these errors
being, in some sense, in opposition, the optimal 
\add{}{$\alpha$-LES}
 solution is found
at a lower resolution than the grid-independent solution.  Indeed, we
see by examining Fig. \ref{FIG:LES_ERR} (a) that for a given filter
the combination of truncation error and the \lansa solution is a
better approximation to the DNS.  For fixed resolution,
Fig. \ref{FIG:LES_ERR} {(b)}, the optimal value for $\alpha$ is not
zero but has some finite value.  This local minimum error shown in the
figure keeps $\alpha$ from going to zero ($k_\alpha\rightarrow\infty$)
in dynamical models \cite{ZM04}.  We note, also, that the error is low for
a finite range of $N$ and $k_\alpha$ near the minimum, indicating that an 
\add{}{$\alpha$-LES} solution may perform well for a range of parameters near the
optimal ones.  We find the resolution for an optimal $\alpha$-LES is also
predicted by Eq. (\ref{EQ:THEORY}) (with $C \approx 47$ for
$Re\approx670$, or $C_0\approx1.8$).  That is, optimal \add{}{$\alpha$-}LES resolution
is just below that for grid-independent \lansa solutions.  Having
demonstrated the predictability of the resolution for grid-independent
\lansa and of \lansa LES given a Reynolds number and a filter, in the
following section we seek to determine sufficient conditions on the
free parameter $\alpha$ for \add{}{\lansa to be} a successful LES.

\subsection{Free parameter $\alpha$ and quality of the $\alpha$-LES}
\label{LES_QUALITY}

\begin{figure}[htbp]
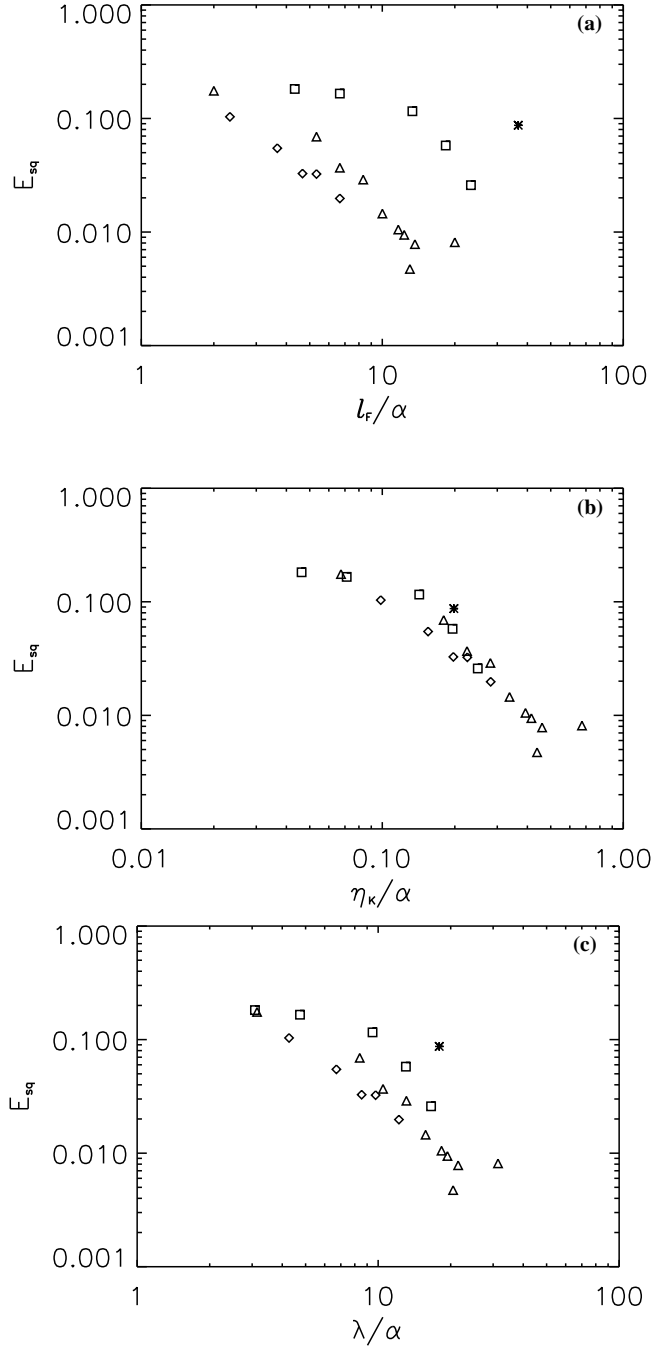
\thesis{\begin{center}\leavevmode}
\vspace{-3mm}
  \includegraphics[width=\paper{8.95}\thesis{8}cm]{\thesis{scaling/}fig11a}
\vspace{-6mm} 
  \includegraphics[width=\paper{8.95}\thesis{8}cm]{\thesis{scaling/}fig11b}
\vspace{-9mm} 
  \includegraphics[width=\paper{8.95}\thesis{8}cm]{\thesis{scaling/}fig11c} 
\vspace{3mm} 
  \caption[Plot of errors of grid-independent solutions compared to
DNS versus ratio of $\alpha$ to the forcing scale, Taylor scale, and dissipation scale.] {Plot of errors, Eq. (\ref{EQ:ERR_2SC}), of
grid-independent solutions compared to DNS.  Asterisks are for $Re
\approx 8300$, squares for $Re \approx 3300$, triangles for $Re
\approx 670$, and diamonds for $Re \approx 500$.  The single
right-most triangle in all plots corresponds to a value of $\alpha$ in
the dissipative range ($k_\alpha=60$).  The norm we employ to measure
the error, Eq. (\ref{EQ:ERR_2SC}), is no longer a good norm when
dissipative scales are considered.  {\bf(a)} Errors versus
$l_F/\alpha$.  No clear correlation between LES quality and the ratio
of the forcing scale to $\alpha$ holds independently of \Res. {\bf(b)}
Errors versus ratio of dissipative scale, $\eta_K$, to $\alpha$.  The
quality of the LES appears to be closely tied to this ratio. {\bf(c)}
Errors versus ratio of Taylor wavenumber, $\lambda$, to $\alpha$.  The
$Re\approx8300$ experiment (asterisk) indicates that the quality of
the {$\alpha$-}LES is not tied to the Taylor scale.  }
  \label{FIG:ERROR1}
\thesis{\end{center}}
\end{figure}

\begin{figure}[htbp]
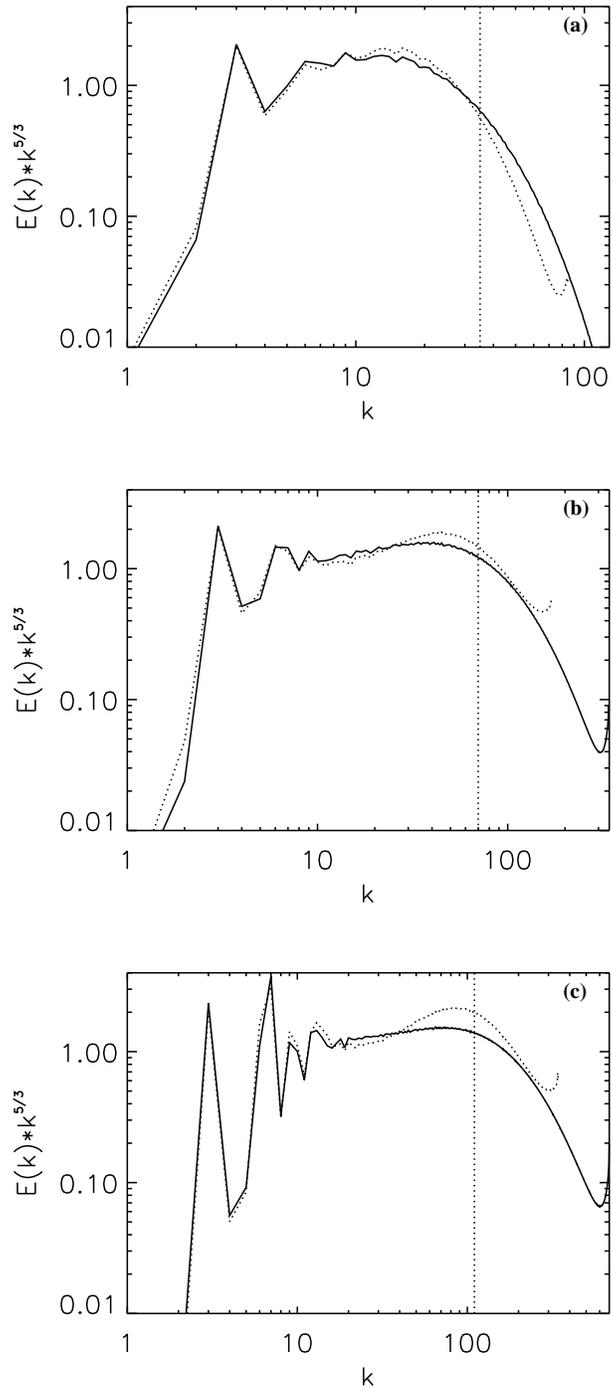
\thesis{\begin{center}\leavevmode}
  \includegraphics[width=\paper{8.95}\thesis{8.5}cm]{\thesis{scaling/}fig12a}
  \includegraphics[width=\paper{8.95}\thesis{8.5}cm]{\thesis{scaling/}fig12b}
  \includegraphics[width=\paper{8.95}\thesis{8.5}cm]{\thesis{scaling/}fig12c}
  \caption[Compensated averaged grid-independent energy spectra for
DNS and \lansa holding the ratio of Taylor scale to $\alpha$
nearly constant] {Compensated averaged grid-independent energy spectra for
DNS (solid) and \lansa (dotted) holding the ratio of Taylor scale
$\lambda$ to $\alpha$ nearly constant.  Vertical dotted lines indicate
$k_\alpha$.  {\bf(a)} $Re\approx670$ and $k_\alpha = 35$
($\lambda/\alpha=18$).  {\bf(b)} $Re\approx3300$ and $k_\alpha = 70$
($\lambda/\alpha=17$). {\bf(c)} $Re\approx8300$ and $k_\alpha = 110$
($\lambda/\alpha=17$).  We see that the maximum deviation from the DNS
increases with $Re$.  This is due to the greater distance between
$\alpha$ and the dissipative scale $\eta_K$.  (Note that scales larger
than $k=3$ are affected by numerical truncation issues.) }
  \label{FIG:ERROR2}
\thesis{\end{center}}
\end{figure}

In this section, we make an analysis of the LES potential of \lansa by
considering only the grid-independent \lansa solutions identified
using Eq. (\ref{EQ:THEORY}).  Note that from the results discussed in
the previous section, we expect \lansa optimal grid-dependent \neu{}{$\alpha-$}LES
approximations to have better performance.  In the limit of $\alpha$
going to zero, \lansa Eq. (\ref{eq:lansSC}) recovers the Navier-Stokes
equations, Eqs. (\ref{eq:navier-stokes}), but the question we address
now is how small must $\alpha$ be for \lansa solutions to be good
approximations to Navier-Stokes solutions.  There are several length
scales that $\alpha$ could be related to: the forcing scale $l_F$, the
integral scale $L = 2\pi\int_0^\infty E(k)k^{-1} dk / \int_0^\infty
E(k) dk$, the Taylor microscale $\lambda$, or the Kolmogorov
dissipation scale $\eta_K$.  Plots of the mean square spectral errors to DNS
(see Eq. (\ref{EQ:ERR_2SC})) versus these scales are shown in
Fig. \ref{FIG:ERROR1}.  While the general trend of errors decreasing
with $\alpha$ is apparent in all cases, in Fig. \ref{FIG:ERROR1}(a) we
see a large difference between errors at varying Reynolds numbers and
similar ratios of $\alpha$ to the forcing scale, $l_F$.  For a linear
least-squares fit, the goodness-of-fit, $\chi^2 \equiv \sum
(E_{sq}^{actual}-E_{sq}^{fit})^2$, was found to be $\chi^2 =
6.2\times10^{-2}$.  The errors for $Re\approx3300$ are much larger than
for the same ratio $l_F/\alpha$ as results at both $Re\approx500$ and
$Re\approx670$.  This is also the case for the integral scale.
However, the quality of the \add{}{$\alpha$-}LES appears to be closely tied to the
ratio of $\alpha$ to the Kolmogorov dissipation scale.  In
Fig. \ref{FIG:ERROR1}(b) the errors are plotted versus the ratio of
the dissipation scale, $\eta_K$, to $\alpha$.  We see a very strong
dependence ($\chi^2 = 2.5\times10^{-2}$) between errors for several
runs with four different Reynolds numbers indicating that the quality
of the \lansa LES approximation is a function of the ratio of $\alpha$
to the dissipative scale.  Finally, in Fig. \ref{FIG:ERROR1}(c) the
errors are plotted versus the ratio of the Taylor Scale, $\lambda$, to
$\alpha$.  We find $\chi^2 = 3.1\times10^{-2}$ for a linear
least-squares fit.  We note that a single experiment conducted at
$Re\approx8300$ (the asterisks) confirms that the maximal value of
$\alpha$ is tied to the dissipation scale and not the Taylor scale.  This is more clearly demonstrated in Fig. \ref{FIG:ERROR2} where we
plot compensated energy spectra for a nearly constant ratio
$\lambda/\alpha$ at three Reynolds numbers.  We see that the maximum
deviation from the DNS spectrum increases with $Re$.  As
$\lambda/\alpha$ is the same in all cases, the optimal $\alpha$ is not
dependent on the Taylor scale.

\begin{figure}[htbp]
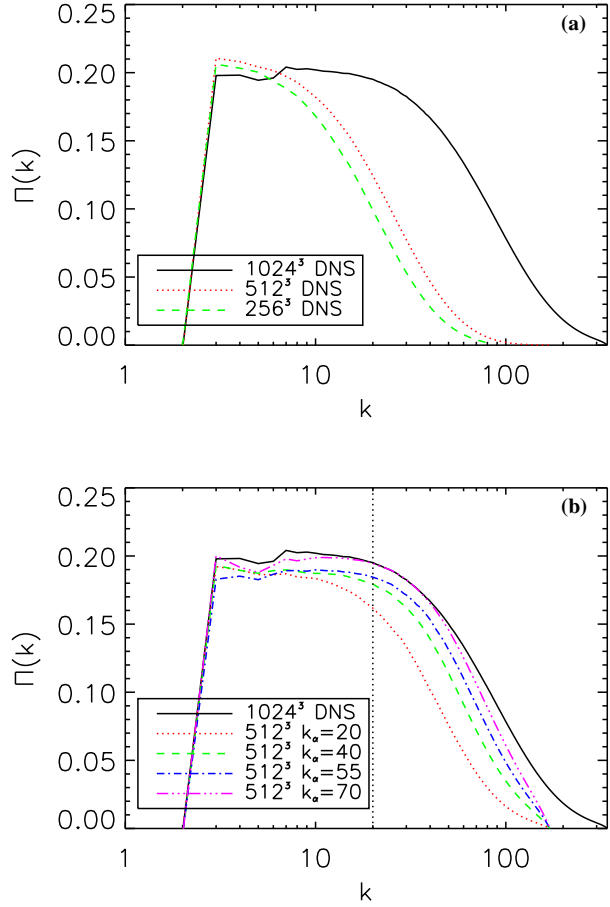
\thesis{\begin{center}\leavevmode}
  \includegraphics[width=\paper{8.95}\thesis{12.5}cm]{\thesis{scaling/}fig13a}
  \includegraphics[width=\paper{8.95}\thesis{12.5}cm]{\thesis{scaling/}fig13b}
  \caption[Energy flux for three DNS with $Re\approx3300$,
$Re\approx670$, and $Re\approx500$ and Energy flux at $Re\approx3300$
for both DNS and $\alpha-$model runs] {\paper{(Color online.)}  {\bf (a)} Energy flux,
Eq. (\ref{EQ:FLUX}), for three DNS with $Re\approx3300$ (black,
solid), $Re\approx670$ (red, dotted), and $Re\approx500$ (green,
dashed).  No inertial range is discernible on the flux functions
except for the highest Reynolds number case.  The initial plateau
followed by a bump and another plateau (for the case at the highest
Reynolds number) is a result of the forcing employed.  {\bf (b)}
Energy flux at $Re\approx3300$ for both DNS and $\alpha-$model runs;
DNS is the black, solid line.  See inset for \lansa parameters.
{\lansa gives} a reduced flux which is linked to the significant
pile-up of energy at high wavenumber as visible in the energy spectrum
(see Fig \ref{FIG:BAD_ALPHA_70}).  Plots of $\varepsilon_\alpha$
versus $t$ (not shown) also show that flux decreases (on average, at
long times) with increasing $\alpha$.  }
  \label{FIG:Flux}
\thesis{\end{center}}
\end{figure}

These findings were not accessible at lower Reynolds numbers due to
inadequate separation of scales.  For example, we give in
Fig. \ref{FIG:Flux}(a) spectral flux for DNS at $Re \approx 500$, $670$,
and $3300$ respectively.
\thesis{\input{derivations/Transfer.tex}}
  We define the kinetic energy transfer
function, $T(k)$, in Fourier space as $T(k) = -\int
\hat{\vec{v}}_k\cdot \widehat{(\boldsymbol{\omega}\times\vec{v})} dV$, where
$\hat{(\cdot)}$ represents the Fourier transform.  For \lansa we have
$T_\alpha(k) = -\int \tilde{\vec{u}}_k\cdot
\widehat{\left(\boldsymbol{\omega}\times\vec{u}\right)} dV$ where
$\boldsymbol{\omega} = \vec{\nabla} \times \vec{v}$.  The flux is defined as
usual from the transfer function as
\begin{equation}
\Pi_{(\alpha)}(k) = \int_0^kT_{(\alpha)}({k^{'}}) dk^{'}.
\label{EQ:FLUX}
\end{equation}
Only $Re\approx3300$ (and $Re\approx8300$ not pictured here)
demonstrates a range of nearly constant flux (a well-defined inertial
range) before the dissipation scales.  Following the scaling arguments
in Ref. \cite{FHT01}, one effect of the $\alpha-$model is to increase
the time scale for the cascade of energy to small scales.  This
reduces the flux as $\alpha$ increases ($k_\alpha$ decreases)
\neu{}{as do the hypothesized ``rigid bodies;'' this} can
be seen in Fig. \ref{FIG:Flux}(b).  (Note that in DNS at high
resolution, 80\% of the flux is from local interactions which is
strongly suppressed at scales smaller than $\alpha$ \cite{AMP05b}.)
As dissipation dominates the flux for low and moderate Reynolds
number, the reduced flux of the $\alpha-$model has little consequence
for these simulations.  With a substantial inertial range, however,
this reduced flux results in a pile-up of energy for scales larger
than the dissipative scale and the spectrum approaches the $k^{1}$
spectrum discussed in Section \ref{SCALING}.  As a consequence of the
integral conservation of energy ($E_\alpha = \int
\vec{u}\cdot\vec{v}$) there is a corresponding decrease of energy at
large scales.  Consequently, as the inertial range increases, $\alpha$
must be moved to smaller and smaller scales in order for \lansa not to
alter scales larger than $\alpha$.  In summary, the $\alpha-$model's
reduced flux of energy to small scales is more crucial when the
dissipation scale is farther away from $\alpha$.

\subsection{Numerical savings from employing \lans}

If $\alpha$ must be directly proportional to the Kolmogorov
dissipation scale, we can estimate the LES computational
savings of the \lansa model.  For the Navier-Stokes equations we have
$\mbox{\sl dof}_{NS} \propto Re^{9/4}$ and, as we verified in Section
\ref{GRIDINDEP}, for \lansa we have $\mbox{\sl dof}_\alpha = C_0^3 k_\alpha
Re^{3/2}/27$.  If $k_\alpha$ is directly proportional to the
Navier-Stokes dissipation wavenumber, $k_\eta$, we arrive at
\begin{equation}
k_\alpha \approx \frac{1}{4} k_\eta \propto Re^{3/4},
\end{equation}
and, consequently,
\begin{equation}
\mbox{\sl dof}_\alpha^{LES} \propto Re^{9/4}.
\label{eq:nogain}
\end{equation}
Note that for free $\alpha$, $\mbox{\sl dof}_\alpha$ ({\sl dof} of
\lans) is much smaller than $\mbox{\sl dof}_{NS}$.  But, to obtain an
optimal LES, $\alpha$ is tied to $k_\eta$; then the resolution
requirements ($\mbox{\sl dof}_\alpha^{LES}$) are different and the
decrease in necessary computational resolution from employing \lansa
is fixed. \neu{}{In fact, for the forcing and boundary conditions employed,
we find
\begin{equation}
\mbox{\sl dof}_\alpha^{LES} \approx \frac{1}{12}
\mbox{\sl dof}_{NS}.
\end{equation}}
We note that Eq. (\ref{eq:nogain}) is consistent with
theoretical predictions given in Ref. \cite{GH06b}.  Other LES such as
the similarity model \cite{B80} and the nonlinear (or gradient) model
\cite{L74,CFR79} have also exhibited the characteristic that they
achieve only moderate reductions in resolution and are, therefore,
frequently used in mixed models with a Smagorinsky term (see, e.g.,
\cite{MK00}).  That such additional terms will be required for \lansa
to reproduce the energy spectrum of high $Re$ flows, may not be a
significant factor in its usability.  Note that the usual addition of
extra dissipative subgrid-stress terms (as in the Smagorinsky model)
also introduces a stronger dependence of the system of equations with
the spatial resolution, since the filter width in such models is often
associated to the maximum wavenumber in the box, $k_{max}$.  In that
case, it can make more sense to use grid-dependent solutions of \lansa
(discussed at the beginning of Section \ref{LES}) which give an
optimal \lansa LES, and can as a result give an extra gain in the
computational costs.

\add{}{
We also conclude that, with the scale $\alpha$ being tied to the dissipation scale $\eta_K$, the model \lansa behaves more like a quasi-DNS by opposition to a traditional LES. Note however that a factor of $\approx2.3$ in resolution gain translates into a factor 27 in CPU and a factor 12 in memory savings, still a substantial gain.}

\section{\lansa at very high Reynolds number}
\label{RESULTS}

\begin{figure}[htbp]\thesis{\begin{center}\leavevmode}
  \includegraphics[width=\paper{8.95}\thesis{12.5}cm]{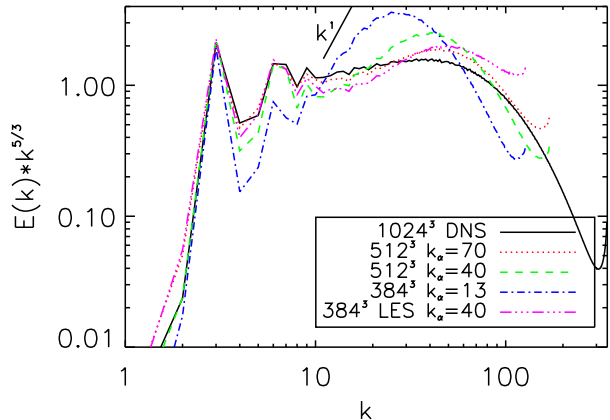}
  \caption[Averaged compensated energy spectra, $Re\approx3300$]
        {\paper{(Color online.)}  Compensated energy spectra averaged over
        $t\in[8,9]$, $Re\approx3300$.  DNS is the solid black line and
        grid-independent \lansa solutions are shown as (red online)
        dotted ($k_\alpha=70$), (green) dashed ($k_\alpha=40$), and
        (blue) dash-dotted ($k_\alpha=13$) lines, respectively.  A
        single \lansa LES is shown as a (pink) dash-triple-dotted line
        ($k_\alpha=40$, $N=384$).  The LES is seen to better
        approximate the DNS spectrum than the grid-independent
        solution for the same value of $\alpha$ ($2\pi/40$).  As
        $\alpha$ is increased the energy spectrum approaches the $k^1$
        spectrum discussed in Section \ref{SUBDOMINANT}.}
  \label{FIG:BAD_ALPHA_70}
\thesis{\end{center}}
\end{figure}

In this section, we compare and contrast \lansa and Navier-Stokes
solutions at high Reynolds number.  Using results of previous sections
for optimal resolution and the necessary value of $\alpha$ to
approximate DNS, we now evaluate both grid-independent \lansa
solutions and a single \lansa LES for a highly turbulent flow
($Re\approx3300$, $R_\lambda\approx790$).  We calculate
grid-independent solutions for $k_\alpha = 70$ ($N=512$), for
$k_\alpha = 40$ ($N=512$), and for $k_\alpha = 13$ ($N=384$).  A
\lansa LES solution is computed for $k_\alpha = 40$ ($N=384$).  Averaged compensated energy spectra are shown in
Fig. \ref{FIG:BAD_ALPHA_70}.  We see that the optimal \lansa LES is a
better approximation of the DNS spectra than the grid-independent
\lansa for the same value of $\alpha$ ($2\pi/40$).  We also see that
if $\alpha$ is increased further, the energy spectrum approaches the
$k^1$ spectrum discussed in Section \ref{SUBDOMINANT}.

\begin {figure}[htbp]
\thesis{\begin{center}\leavevmode}
  \includegraphics[width=\paper{8.95}cm]{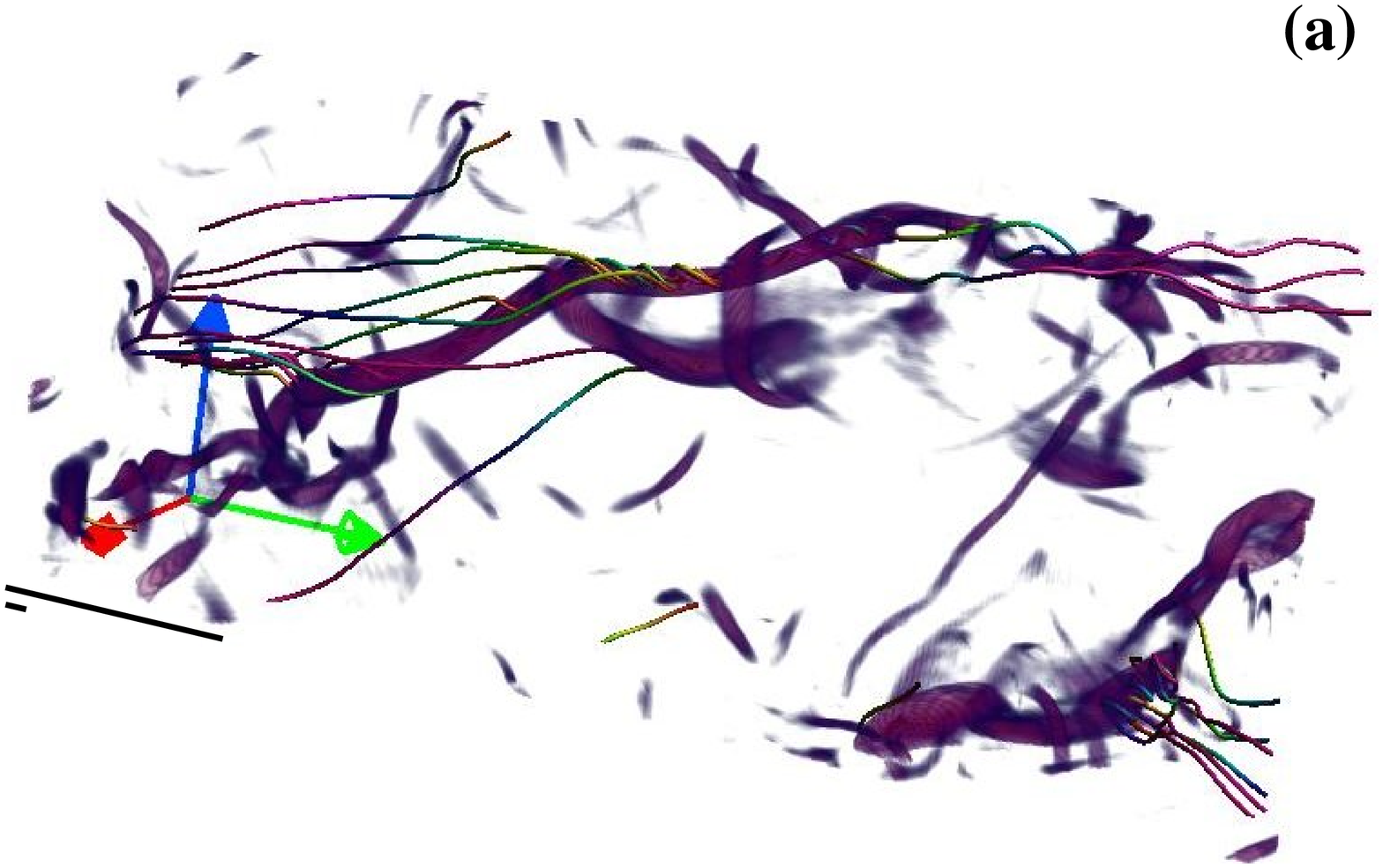}
  \includegraphics[width=\paper{8.95}cm]{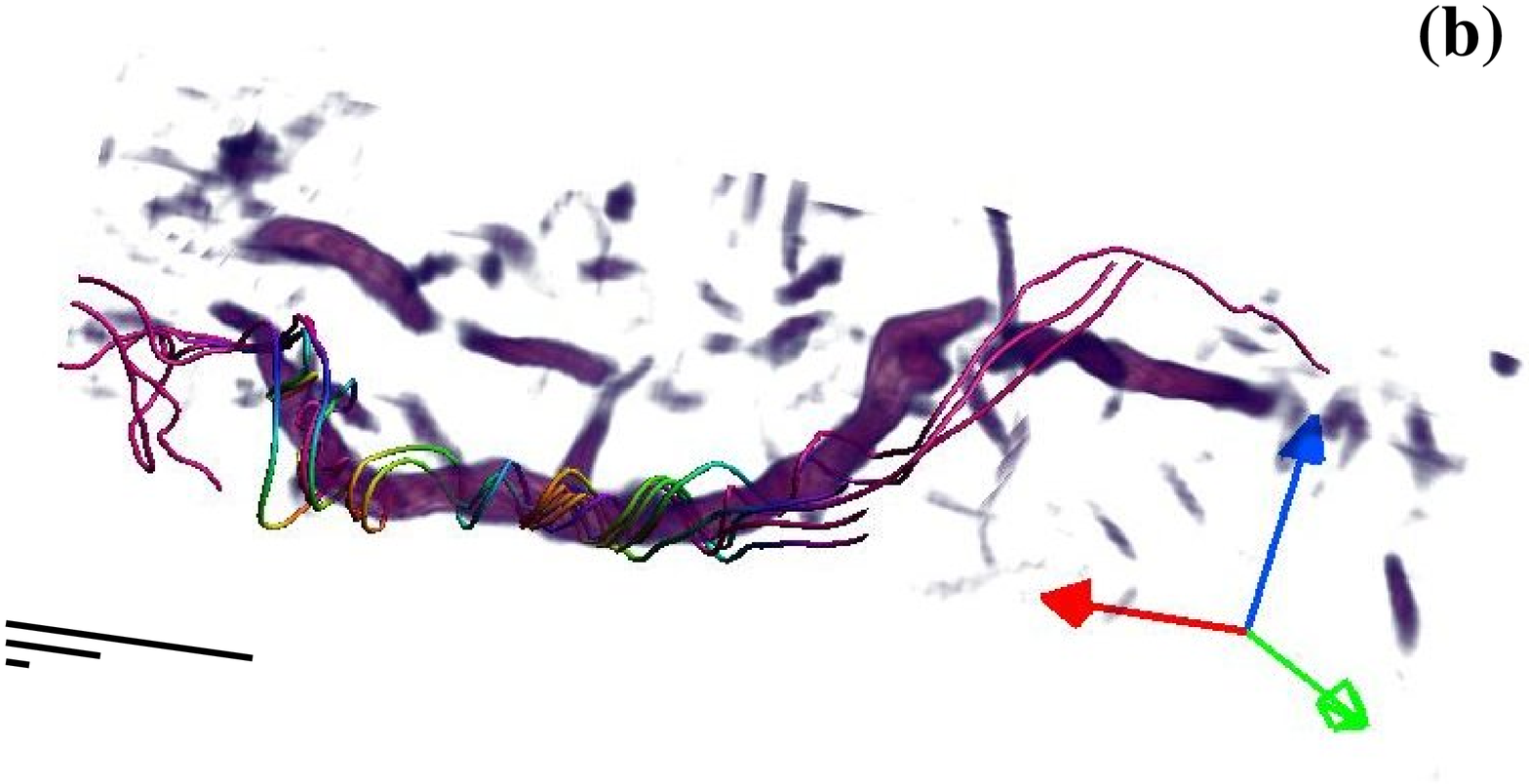}
  \includegraphics[width=\paper{8.95}cm]{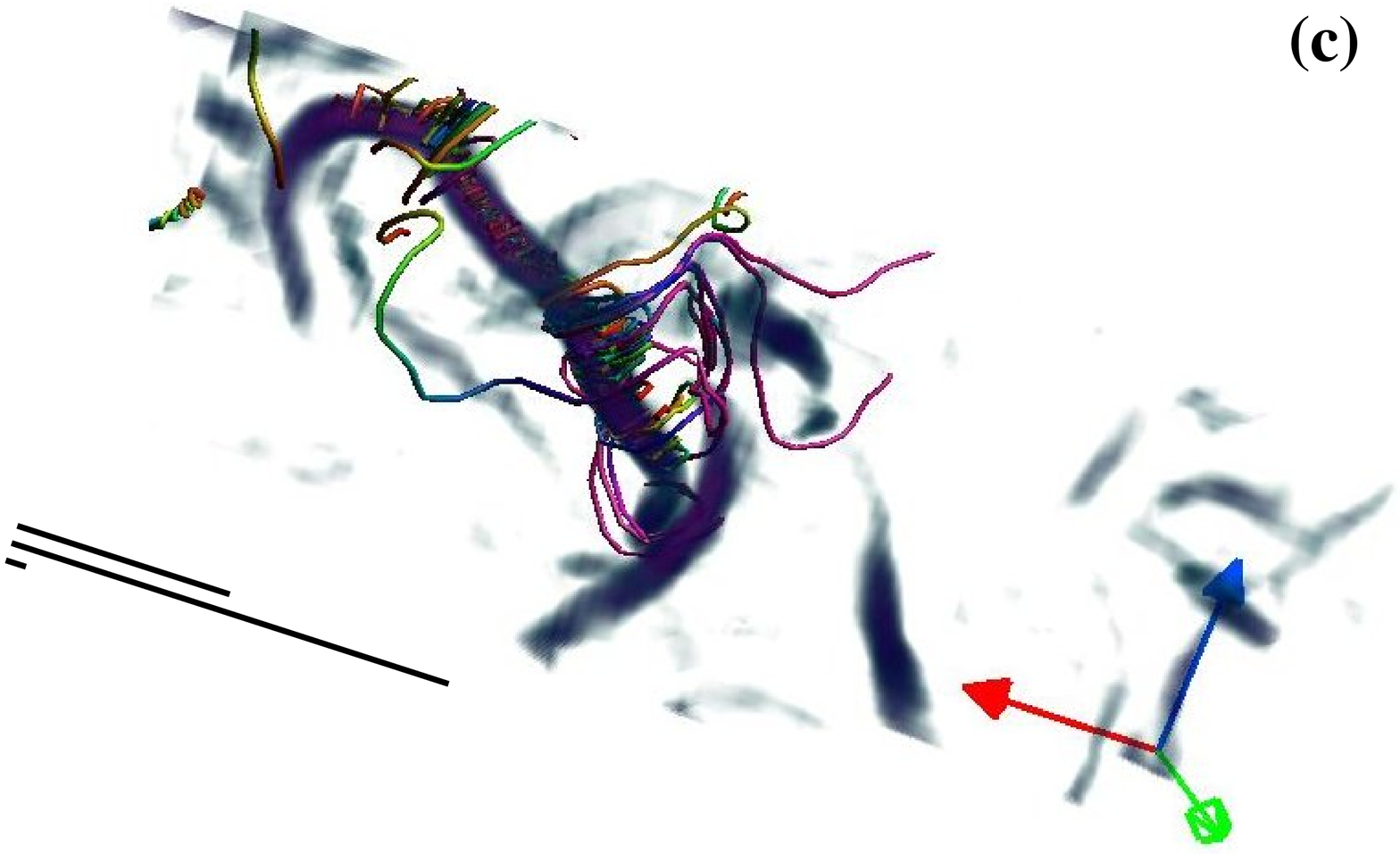}
  \caption [3D volume rendering of enstrophy density]{\paper{(Color online.)}  Rendering of enstrophy density $\omega^2$
($\boldsymbol{\omega}\cdot\boldsymbol{\bar{\omega}}$ for \lans).  Due to the late
time depicted here ($t=9$, longer than a Lyapunov time) there can be
no point-by-point comparison between the simulations.  Instead,
regions with approximately the same dimensions are selected around
vortex tubes.  Velocity $\vec{v}$ field lines are also shown
illustrating the helical nature of the tubes which is seen to be
captured by \lans.  {\bf(a)} DNS.  The thick bars represent, from top to
bottom, the Taylor scale $\lambda$ and the dissipative scale $\eta_K$,
respectively.  For \lansa results the scale $\alpha$ is depicted
between these two.  {\bf(b)} $k_\alpha=70$, $N=512$. {\bf(c)}
$k_\alpha=13$, $N=384$.  We see that, for large values of $\alpha$, the
vortex tubes become shorter and somewhat thicker.  }
  \label{fig:VAPOR_TUBE}
\thesis{\end{center}}
\end{figure}

Fig \ref{fig:VAPOR_TUBE} is a perspective volume rendering of the enstrophy density
$\omega^2$ ($\boldsymbol{\omega}\cdot\boldsymbol{\bar{\omega}}$ for \lans) for the
DNS, $k_\alpha=70$ \lans, and $k_\alpha=13$ \lans.  Due to the late
time depicted here ($t=9$, longer than a Lyapunov time) there can be
no point-by-point comparison between the simulations.  However, we
note that the helical structure of the vortex tubes is preserved by
the $\alpha-$model but that the tubes themselves are shorter and
somewhat thicker for large values of $\alpha$.  As was noted for
moderate Reynolds numbers, this is due to \lansa suppressing vortex
stretching dynamics without changing its qualitative features
\cite{CHM+99}.  This is in contrast to 2D \lansa where the vorticity
structures are seen to get thinner as $\alpha$ increases
\cite{LKT+07}.  This could also be related to the scaling differences
between 2D and 3D \lans.  It has been claimed that the development of
helical structures in turbulent flows can lead to the depletion of
nonlinearity and the quenching of local interactions \cite{MT92,T01}.
The depletion of energy transfer due to local interactions at some
cutoff in wavenumber is also believed to bring about the bottleneck
effect \cite{HSL+82,LMG95,MCD+97,MAP06}.  Consistent with these
results, in 2D \lansa (where the vorticity structures are more fine
than Navier-Stokes) the spectrum is steeper and in 3D \lansa (where
the vorticity structures are shorter but fatter than Navier-Stokes)
the spectrum is shallower.

\begin{figure}[htbp]\thesis{\begin{center}\leavevmode}
\includegraphics[width=\paper{8.95}\thesis{12.5}cm]{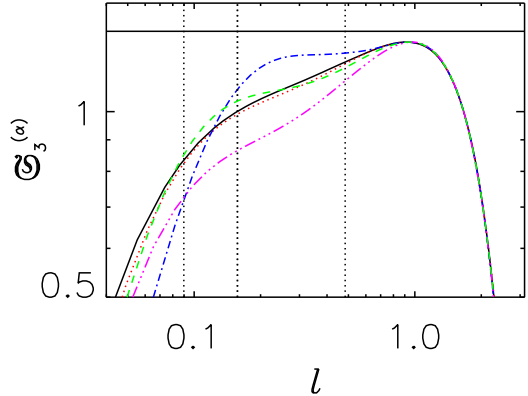}
\caption [Compensated 3rd-order structure function versus length
$l$]{\paper{(Color online.)}  Compensated 3rd-order structure function
versus length $l$ (a horizontal line scales with $l$).  Structure
functions corresponding to the K\'arm\'an-Howarth theorem are depicted
($\mathfrak S_3$ for DNS, $\mathfrak S_3^\alpha\equiv\langle (\delta u)^2\delta v\rangle$ for
\lans).  Labels are as in Fig. \ref{FIG:BAD_ALPHA_70}.  The dotted
vertical lines indicate the various $\alpha$'s.  A small inertial
range for the DNS near $l=1$ is reproduced by \lans.  The largest
$\alpha$ ($2\pi/13$) exhibits a second inertial range at scales just
smaller than $\alpha$ ($\langle (\delta u)^2\delta v\rangle\sim l$ is
consistent with Eq. (\ref{EQ:LCUBElans})).  }
\label{fig:Struct3}
\thesis{\end{center}}
\end{figure}

\begin{figure}[htbp]\thesis{\begin{center}\leavevmode}
\paper{\includegraphics[width=8.95cm]{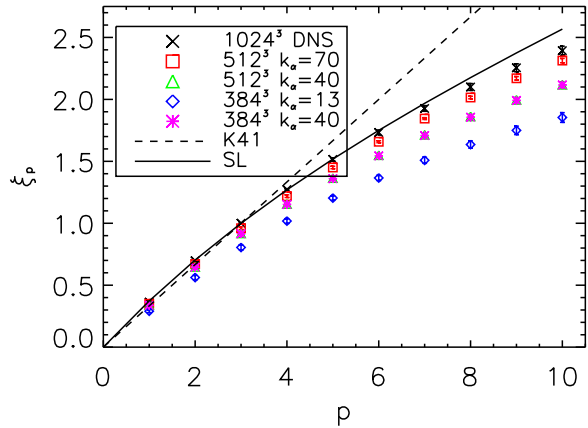}}
\thesis{\includegraphics[width=12.5cm]{eps/fig_exponents}}
\thesis{\includegraphics[width=12.5cm]{eps/fig_exponents_norm}}
\caption [Structure function scaling exponent versus order]{\paper{(Color online.)}  \thesis{{\bf(a)}} Structure function scaling exponent $\xi_p$
versus order $p$.  Black X's are shown for the DNS.  Grid-independent
\lansa are shown as (red\paper{ online}) boxes ($k_\alpha=70$), as (green)
triangles ($k_\alpha=40$), as (blue) diamonds ($k_\alpha=13$).  \lansa
LES ($k_\alpha=40$, $N=384$) is shown as (pink) asterisks.  The dashed
line indicates K41 scaling and the solid line the She-L\'ev\^eque (SL)
formula \cite{SL94}. \thesis{{\bf(b)} Scaling exponents normalized to
compare deviations from linearity.  $k_\alpha=13$ is more
intermittent than the DNS.}}
\label{fig:Scaling}
\thesis{\end{center}}
\end{figure}

Figure \ref{fig:Struct3} shows the third-order (mixed) structure
functions corresponding to the K\'arm\'an-Howarth theorems versus
length $l$.  For the DNS, we show $\mathfrak S_3\equiv\langle \delta v^3\rangle$ and $\mathfrak S_3^\alpha\equiv\langle (\delta
u)^2\delta v\rangle$ for \lans. The dotted vertical lines indicate the
various $\alpha$'s.  A small inertial range for the DNS near $l=1$ is
reproduced by all \lansa results.  The largest $\alpha$ ($2\pi/13$)
exhibits a second inertial range at scales just smaller than $\alpha$
($\langle (\delta u)^2\delta v\rangle\sim l$ is consistent with
Eq. (\ref{EQ:LCUBElans})).  We note this is the first demonstration of
third-order structure functions in \lansa consistent with a K41
inertial range followed by an $\alpha$ inertial range and finally a
dissipative range. Next, we observe the scaling of the longitudinal
structure functions,
\begin{equation}
{S}_p(l) \equiv \langle|\delta v_{\|}|^p\rangle,
\end{equation}
where we again replace the $H^1_\alpha$ norm for the $L^2$ norm in the
case of \lans,
\begin{equation}
{S}_p^\alpha(l) \equiv \langle|\delta u_{\|}\delta v_{\|}|^{p/2}\rangle.
\end{equation}
We utilize the extended self-similarity (ESS) hypothesis
\cite{BCB+93,BCT+93,BBC+96} which proposes the scaling
\begin{equation}
{S}_p(l) \propto {S}_3(l)^{\xi_p}
\end{equation}
or, for \lans,
\begin{equation}
{S}_p^\alpha(l) \propto \langle (\delta
u)^2\delta v\rangle^{\xi_p}.
\end{equation}
We display our results in Fig. \ref{fig:Scaling}\thesis{(a)}.  We note that for \lans,
the third-order exponent is not\add{}{ equal to unity, contrary to} the Navier-Stokes case.  The
K\'arm\'an-Howarth theorem implies $\langle (\delta u)^2\delta
v\rangle\sim l$, not ${S}_3^\alpha(l)\sim l$.  We measured the
deviation from linearity for each experiment \paper{(not depicted here)}\thesis{(see Fig. \ref{fig:Scaling}(b))} and
found that \lansa becomes more intermittent as $\alpha$ increases
($k_\alpha=13$ is slightly more intermittent than the DNS).  As
artificially dropping local small-scale interactions gives enhanced
intermittency \cite{LDN01,DLN+04}, this increased intermittency is the
expected result of \lansa reducing interactions at scales smaller than
$\alpha$.  \neu{}{We note, however, that even with such a large
filter, \lansa is a good approximation to the intermittency properties
of the DNS.  This is surprising given its energy spectrum and reduced
flux in the inertial range.} \add{}{It is probably linked to the fact that \lansa preserves global properties
(in an $H^1$ sense) of the Navier-Stokes equations and that these properties are important to the dynamics of small scales as measured by high-order structure functions.}

\section{Conclusions}
\label{SEC:CONCL}

We computed solutions of the Lagrangian-Averaged Navier-Stokes
$\alpha-$model \thesis{\\}(\lans) in three dimensions for significantly higher
Reynolds numbers (up to $Re \approx 8300$) than have previously been
accomplished and performed numerous forced turbulence simulations of
\lansa to study their equilibrium states.  The results were compared
to DNS for $Re\approx500, 670,$ $3300$, and $8300$
\add{}{, the last performed on a grid of $2048^3$ points}.  
We note that
there are two ways to view the \lansa simulations: as converged or
``grid-independent'' solutions of the \lansa equations or as
large-eddy simulations (\neu{}{$\alpha-$}LES) which include grid effects.  We found a
definite difference between the two approaches in that the
fully-converged grid-independent \lansa is not the best approximation
to a DNS of Navier-Stokes.  Instead, the minimum error is a balance
between truncation errors and the approximation error due to using
\lansa instead of the full Navier-Stokes equations.  Due to these
errors being, in some sense, in opposition, the optimal \add{}{$\alpha$-}LES solution
was found at a lower resolution than the grid-independent solution
(the error was low for a finite range of $N$ and $\alpha$ near the
minimum, indicating that a \lansa viewed as an LES solution may perform well for a range
of parameters).  Unlike the 2D case \cite{LKT+07}, 3D \lansa has been shown to be a
subgrid model (i.e., it reduces the resolution requirements of a given
computation).  This difference between 2D and 3D \lansa indicates that
other $\alpha-$models (as the LAMHD$-\alpha$
Eqs. \cite{PGMP05,PGHM+06} or the BV$-\alpha$ Eqs. \cite{HN03}) may
behave differently and studies of these systems at high
resolution may be required.

We confirm the presence of the theoretically predicted $l^3$ scaling
of the third-order structure function (corresponding to a $k^{-1}$
scaling of the energy spectrum) \cite{FHT01,CHT05,CHO+05} through
its bound on the number of degrees of freedom for \lansa \cite{FHT01},
in the structure functions of the smoothed velocity in simulations
with large $\alpha$, \neu{}{and in the spectrum of specific spatial portions of the flow}.  In so doing, we have validated the predictive
power of the bound $\mbox{\sl dof}_\alpha < \mathcal C
\alpha^{-1}Re^{3/2}$, for the numerical resolution for
grid-independent \lansa solutions and for optimal \lansa LES (with a
separate constant of proportionality).  The great utility of the
prediction is that the single constant can cheaply be determined at
low and moderate Reynolds number and predicts the resolution
requirement for the highest Reynolds numbers attainable.  We further
found no great change in this single constant when employing the
non-helical Taylor-Green or the maximally-helical ABC forcings.

However, the small scale ($k\alpha\gg1$) \lansa spectrum was observed
to be $k^{+1}$.  We attribute this to the frozen-in-turbulence closure
employed in deriving the $\alpha-$model.  For scales smaller than
$\alpha$, portions of the smoothed flow $\vec{u}$ are locked into
``rigid \neu{}{bodies}.''  \neu{}{By ``rigid bodies,'' we mean
    the internal degrees of freedom are frozen and these portions give no
    contribution to the energy cascade.}  This is consistent both with the observed
$k^{+1}$ spectrum and with field increments $\delta u_{\|}$ being
observed to be approximately zero over a large portion (compared to
Navier-Stokes) of the flow.  The turbulent energy cascade occurs in
the space between these ``rigid'' portions. While the $k^{-1}$
\neu{}{portions are} subdominant to the $k^{+1}$ \neu{}{portions} in the energy spectrum, they prevail in the
cascade and hence both the structure functions and the degrees of
freedom of the \lansa attractor.

We find that both of these scalings ($k^{+1}$ and $k^{-1}$) contribute
to a reduction of flux at constant energy (i.e., the dissipation is
reduced as has previously been observed in 2D calculations
\cite{BS01}).  This leads to a shallower (or even growing) energy
spectrum as $\alpha$ increases.  Thus, for \lansa \add{}{viewed as an} LES to reproduce the
Navier-Stokes energy spectrum it is necessary that $\alpha$ be not
much larger than the dissipation scale ($\alpha \lessapprox 4\eta_K$
independent of Reynolds number); 
\add{}{in that sense, it can be considered as a quasi-DNS as opposed to a traditional LES, substantially larger Reynolds numbers being modeled in the latter case, leading to substantially larger gain in resolution}.  As a consequence, the computational
savings of \lansa is fixed and not a function of Reynolds number.
(However, and unlike the 2D case, the 3D $\alpha-$model does give a
computational saving when used as a LES.)  This result {\sl was not
accessible at lower Reynolds numbers due to inadequate separation of
scales}.  However, in one previous study for decaying turbulence with
energy initially mostly at low wavenumbers ($k=3$), it was evident that
as time evolved and energy moved to smaller scales, the resolution
requirements of \lansa increased \cite{MKS+03}.  Other LES such as the
similarity model \cite{B80} and the nonlinear (or gradient) model
\cite{L74,CFR79} have also exhibited the characteristic that
resolution may be decreased only modestly and are, therefore,
frequently used in mixed models with a Smagorinsky term (see e.g.,
\cite{MK00}).  That such additional terms will be required for \lansa
to reproduce the energy spectrum of high $Re$ flows, may not be a
significant factor in its usability.

We compared and contrasted \lansa to a DNS at $Re\approx3300$
considering both structures and high-order statistics such as the
longitudinal structure functions which are related with intermittency.
With an appropriate choice of $\alpha$ we were able to observe a
Navier-Stokes inertial range followed by \lansa inertial range at
scales smaller than $\alpha$.  For this second inertial range we again
observed a $k^{+1}$ energy spectrum.  As $\alpha$ increased, we noted
a change in the aspect ratio of vortex tubes (they became shorter and
fatter).  This can be related to quenching of local small-scale
interactions at scales smaller than $\alpha$ and, thus, to the
shallower spectrum for 3D \lansa
\cite{MT92,T01,HSL+82,LMG95,MCD+97,MAP06}.  Therefore, in 2D \lansa
(where the vorticity structures are more fine than Navier-Stokes) the
spectrum is steeper \cite{LKT+07} and in 3D \lansa (where the
vorticity structures are shorter but fatter than Navier-Stokes) the
spectrum is shallower.  Finally, an examination of the longitudinal
structure functions indicate that intermittency is increased as the
parameter $\alpha$ is increased consistent with the suppression of
local small-scale interactions at scales smaller than $\alpha$
\cite{LDN01,DLN+04}.

The elimination of the faster and faster interactions among smaller
and smaller scales through the modified nonlinearity in \lansa
(together with the discrepancy between its solutions and Navier-Stokes
solutions) highlights the importance of these interactions down to
scales only slightly larger than the dissipative scale.  That is, by
removing these interactions anywhere in the inertial range (e.g.,
$\alpha \gtrapprox 4\eta_K$), the resulting energy spectrum was found
to differ from the DNS at scales larger than $\alpha$.  \neu{}{The
intermittency properties of the DNS, however, were well reproduced
even with large filters.  Noting this, if \lans's $k^1$ energy
spectrum is not important for a given application, much greater
reductions in resolution can be achieved.}  Future work should address
whether this may be remedied in a \lansa LES by the inclusion of
another (dissipative) model for these interactions, or (in the case of
magneto-hydrodynamics \cite{PGMP05,PGHM+06} \neu{}{whether this problem} is less significant 
\neu{}{because of }
the presence of greater spectral nonlocality \neu{}{\cite{AMP05a,MAP05,AMP06}.}  The effect of \lansa on
the detailed scale-by-scale energy transfer should also be
investigated as our results indicate that a model for local
small-scale interactions would improve the $\alpha-$model.  Another
direction \neu{}{of future research is to explore other} reduced \lansa models, Clark$-\alpha$ and
Leray-$\alpha$, which break the frozen-in-turbulence closure and,
also, the conservation of circulation.
\thesis{This we do in the following chapter.}
\add{}{Finally, note that because of its greater mathematical tractability, \lansa possibly allows for a better understanding of multi-scale interactions in turbulent flows thus modeled; therefore, detailed studies such as the one presented here may, {\it in fine}, allow for a better understanding of turbulence itself.}


\begin{acknowledgments}
Computer time was provided by NCAR and by the National Science
Foundation Terascale Computing System at the Pittsburgh Supercomputing
Center.  The NSF Grant No. CMG-0327888 at NCAR supported this work in
part and is gratefully acknowledged.  Three-dimensional visualizations
of the flows were done using VAPOR.  The authors would like to express
their gratitude for valuable discussions with Bob Kerr.
\end{acknowledgments}

\begin{spacing}{2}

\end{spacing}

\end{document}